\documentclass[amsmath,amssymb,prb,superscriptaddress,twocolumn,showpacs,longbibliography,floatfix]{revtex4-1}
\usepackage{graphicx}
\usepackage{dcolumn}
\usepackage[colorlinks=true,linkcolor=blue ,citecolor=blue,urlcolor=blue]{hyperref}
\usepackage{multirow}
\usepackage[usenames,dvipsnames]{xcolor}
\usepackage{soul}
\usepackage{braket}
\usepackage{bm}
\usepackage{nicefrac}
\usepackage{siunitx}
\usepackage{color}
\usepackage[utf8]{inputenc}
\usepackage{upgreek}

\newcommand{\HH}{\mathcal{H}}

\newcommand{\NN}{\mathcal{N}}
\newcommand{\ud}{\mathrm{d}}
\newcommand{\iu}{\mathrm{i}}
\newcommand{\Tr}{\mathrm{Tr}}

\newcommand{\VEC}[1]{\mathbf{#1}}
\newcommand{\MR}[1]{\mathrm{#1}}
\newcommand{\MC}[1]{\mathcal{#1}}
\newcommand{\RR}{\mathcal{R}}
\newcommand{\eqand}{\quad \textnormal{and} \quad}
\newcommand{\eqdot}{\quad .}

\newcommand{\onehalf}{\frac{1}{2} }
\newcommand{\dv}[2]{\frac{\textnormal{d}#1}{\textnormal{d}#2}}
\newcommand{\dvn}[3]{\frac{\textnormal{d}^#1 #2}{\textnormal{d}#3^#1}}

\begin{document}

\title{Spin-resolved inelastic electron scattering by spin waves in noncollinear magnets}

\author{Flaviano Jos\'e dos Santos}\email{f.dos.santos@fz-juelich.de}
\author{Manuel dos Santos Dias}
\author{Filipe Souza Mendes Guimarães}
\author{Juba Bouaziz}
\author{Samir Lounis}\email{s.lounis@fz-juelich.de}
\affiliation{Peter Gr\"{u}nberg Institut and Institute for Advanced Simulation, Forschungszentrum J\"{u}lich \& JARA, D-52425 J\"{u}lich, Germany}

\date{\today}

\begin{abstract}
\noindent
Topological non-collinear magnetic phases of matter are at the heart of many proposals for future information nanotechnology, with novel device concepts based on ultra-thin films and nanowires.
Their operation requires understanding and control of the underlying dynamics, including excitations such as spin-waves.
So far, no experimental technique has attempted to probe large wave-vector spin-waves in non-collinear low-dimensional systems.
In this work, we explain how inelastic electron scattering, being suitable for investigations of surfaces and thin films, can detect the collective spin-excitation spectra of non-collinear magnets.
To reveal the particularities of spin-waves in such non-collinear samples, we propose the usage of spin-polarized electron-energy-loss spectroscopy augmented with a spin-analyzer.
With the spin-analyzer detecting the polarization of the scattered electrons, four spin-dependent scattering channels are defined, which allow to filter and select specific spin-wave modes.
We take as examples a topological non-trivial skyrmion lattice, a spin-spiral phase and the conventional ferromagnet.
Then we demonstrate that, counter-intuitively and in contrast to the ferromagnetic case, even non spin-flip processes can generate spin-waves in non-collinear substrates.
The measured dispersion and lifetime of the excitation modes permit to fingerprint the magnetic nature of the substrate.
\end{abstract}



\maketitle

%***********************************************************************
\textbf{Introduction}.
Recently, exquisite magnetic states related to chiral interactions in noncentrosymmetric systems have been discovered and intensively investigated.
They are non-collinear magnetic structures such as skyrmions, anti-skyrmions, magnetic bobbers and spin-spirals~\cite{bogdanov_thermodynamically_1989,rosler_spontaneous_2006,nagaosa_topological_2013,koshibae_theory_2016,rybakov_new_2016,hoffmann_antiskyrmions_2017,nayak_magnetic_2017,zheng_experimental_2017}.
These states arise from the delicate balance of internal and external interactions, such as the magnetic exchange, Dzyaloshinskii-Moriya and magnetic fields, which can trigger topologically non-trivial properties~\cite{bogdanov_chiral_2001,kiselev_chiral_2011,fert_skyrmions_2013,romming_writing_2013}.
Most important for applications is their formation in ultra-thin films, given that they can be tailored by the structure and composition of heterogeneous multilayers~\cite{heinze_spontaneous_2011,romming_writing_2013,dupe_tailoring_2014}. 
Concurrently, spin-waves have been explored for their potential application in spintronic and magnonic devices~\cite{kruglyak_magnonics_2010,lenk_building_2011,pereiro_topological_2014,chisnell_topological_2015,chumak_magnon_2015}.
However, the behavior of spin-waves in these non-collinear systems is only now beginning to be understood~\cite{belitz_theory_2006,janoschek_helimagnon_2010,mochizuki_spin-wave_2012,iwasaki_theory_2014,schwarze_universal_2015,ma_skyrmion-based_2015,Roldan2016,chernyshev_damped_2016,garst_collective_2017,weiler_helimagnon_2017,hog_theory_2017}.

Do spin-waves inherit special properties due to the topology of the magnetic structure, leading to revolutionary applications?
To explore this question we need to understand the manifestation of spin-waves in these novel magnetic phases: how they may be excited, controlled and detected.
Non-collinear magnetic structures intrinsically feature many spin-wave bands (or modes) due to the breaking of translational and rotational symmetries~\cite{toth_linear_2015,Roldan2016}.
However, only a few of them can be excited or detected by a given experimental setup.
Thus, a discussion of spin-wave excitations must go together with the exciting/probing technique.
On the one hand, inelastic neutron-scattering and microwave resonance have been used to investigate collective spin-excitations in bulk chiral helimagnets and two-dimensional skyrmion lattice~\cite{janoschek_helimagnon_2010,mochizuki_spin-wave_2012,schwarze_universal_2015,weiler_helimagnon_2017}.
While the first lacks surface sensitivity, the second is restricted to excitations near the $\Gamma$-point.
On the other hand, inelastic electron scattering has been applied with great success to study spin-waves in ultra-thin films~\cite{plihal_spin_1999,Vollmer2003,tang_large_2007,prokop_magnons_2009,zakeri_asymmetric_2010,zhang_elementary_2011,zakeri_direct_2013,Michel2015,michel_lifetime_2016,dos_santos_first-principles_2017,qin_temperature_2017}, due to the large scattering cross section of the electrons.
However, to the best of our knowledge, it has only been employed for ferromagnets.
The same is true from the theoretical side~\cite{Mills1967,gokhale_inelastic_1992}.

\begin{figure}[t!]
  \centering
  \includegraphics[width=0.48\textwidth,trim={4em 0 15 30},clip=true]{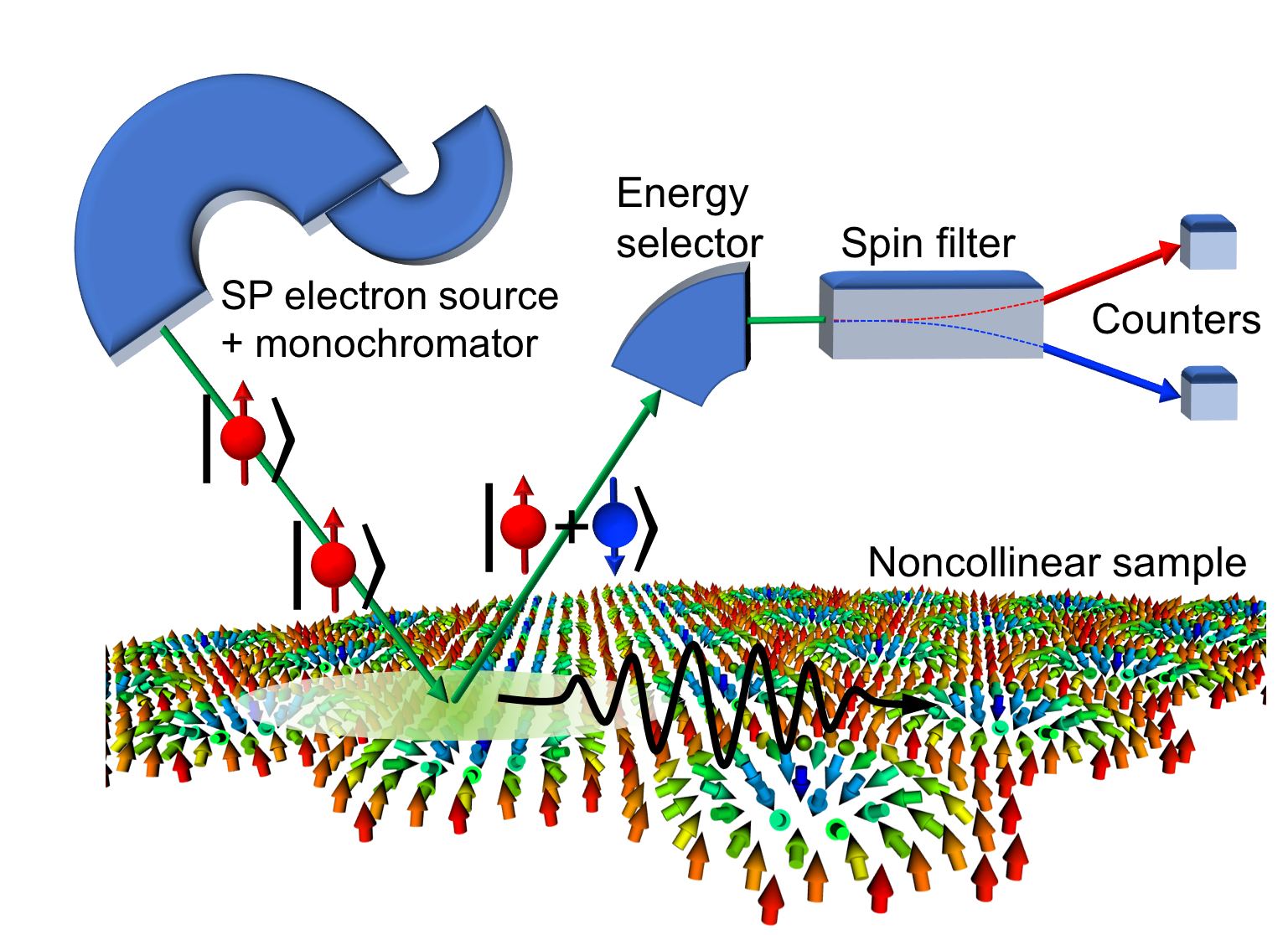}
  \caption{\label{fig:SREELS}
  Schematic picture of spin-resolved electron-energy-loss spectroscopy (SREELS).
  A monochromatic spin-polarized (SP) electron beam is aimed at the surface of a non-collinear magnetic sample.
  The magnetic non-collinearity leads to a mixed spin state of the outgoing electrons.
  These are then collected for spectroscopical analysis, having both their energy and spin characterized.
  }
\end{figure}

In this paper, we provide a quantum description of the inelastic scattering of electrons by spin-waves in non-collinear systems.
We illustrate these developments with two non-collinear phases of an hexagonal monolayer, namely a cycloidal spin-spiral and a skyrmion lattice, contrasting them with the well-known ferromagnetic case.
The spectra were calculated as to be measured by spin-polarized electron-energy-loss spectroscopy augmented with a spin-analyzer, see Fig.~\ref{fig:SREELS}.
We demonstrate that this spin-resolved spectroscopy enlightens the existence of zero net angular momentum spin-waves in non-collinear substrates; and that our proposed scheme permits to filter and select specific spin-wave modes.
We also observe the highly anisotropic dispersion-relation and localization of spin-waves in the helical sample.

%***********************************************************************
\textbf{Theory}.
Let us consider an experimental setup based on spin-polarized electron-energy-loss spectroscopy (SPEELS)~\cite{plihal_spin_1999,Vollmer2003} augmented with a spin-filter for the scattered electrons~\cite{kirschner_direct_1985}, which we call spin-resolved electron-energy-loss spectroscopy (SREELS), see Fig.~\ref{fig:SREELS}.
It consists in preparing a spin-polarized monochromatic electron beam, which then scatters from the first few layers of the sample surface.
Scattered electrons may exchange energy, angular and linear momentum due to creation or annihilation of spin-waves.
By the conservation laws of these quantities, measuring their exchanges informs upon spin-wave states of the magnetic system.
An incoming beam with up or down spin polarization generates outgoing electrons in a quantum superposition of up and down states, due to atomic spin moments not aligned with the beam polarization axis.
Then, by filtering the spin of the outgoing electrons, two non-spin-flip scattering channels, up-up and down-down, and two spin-flip ones, up-down and down-up, are defined.
The meaning of these channels will be discussed later with specific examples.

We consider an incoming (outgoing) beam with energy $E_{\MR{in}}$ ($E_{\MR{out}}$), wavevector $\VEC{k}_{\MR{in}}$ ($\VEC{k}_{\MR{out}}$), and spin projection $s_{\MR{in}}$ ($s_{\MR{out}}$), which interacts with a sample held at zero temperature, i.e., in its ground state.
These variables define the energy absorbed by the sample $\omega = E_{\MR{in}} - E_{\MR{out}}$, and the linear and angular momentum transferred, $\VEC{q} = \VEC{k}_{\MR{in}} - \VEC{k}_{\MR{out}}$ and $m = s_{\MR{in}} - s_{\MR{out}}$, respectively.
There are thus four scattering channels, with angular momentum $m = 0,\pm1$, according to the four possible combinations of $s_{\MR{in}}$ and $s_{\MR{out}}$.

We assume that the electrons couple with the atomic spins via a local exchange interaction $\boldsymbol{\upsigma} \cdot \VEC S_\mu$, where $\mu$ labels the basis atom in the unit cell, $\boldsymbol{\upsigma}$ is the Pauli vector describing the electron spin, and $\VEC S_\mu$ is the vector operator describing the atomic spin.
The details of the derivation can be found in the Appendix~\ref{Apx:inelastic_scattering}, so here we just discuss the outcome.
Starting from the Schr\"odinger equation for the coupled system of electron beam and magnetic sample, time-dependent perturbation theory leads to Fermi's Golden Rule for the transition rate between initial and final electron states:
\begin{equation}\label{eq:transrate}
  \Gamma_m(\VEC{q}, \omega) \propto \sum_{\alpha\beta} \sigma_{s_\MR{in} s_\MR{out}}^\alpha \sigma_{s_\MR{out} s_\MR{in}}^\beta
  \sum_{\mu\nu} e^{\MR{i} \VEC{q} \cdot \VEC{R}_{\mu\nu}} \MC{N}_{\mu\nu}^{\alpha\beta}(\VEC{q},\omega) \quad .
\end{equation}
Here $\alpha,\beta = +,-,z$ and $\sigma^\pm = (\sigma^x \pm \MR{i}\,\sigma^y)/2$, with $z$ being the spin quantization axis of the beam polarization.
The wave nature of the electron beam leads to the Fourier factor connecting the basis atoms in the unit cell ($\VEC{R}_{\mu\nu} = \VEC{R}_\nu - \VEC{R}_\mu$), and is responsible for the unfolding of the spin-wave modes~\cite{dos_santos_first-principles_2017}.
The information about the spin-excitations of the sample is contained in the imaginary part of the spin-spin correlation tensor
\begin{equation}\label{eq:density_tensor22}
\begin{split}
  \mathcal{N}_{\mu\nu}^{\alpha\beta}(\VEC q, \omega) = 
   \sum_{\VEC{k} r}  \delta& \left(\omega - \omega_{r}(\VEC k)\right) \times \\
  & \quad \braket{\tilde 0|S_{\mu}^\alpha(\VEC q)|{\VEC{k} r}} \braket{{\VEC{k} r}|S_{\nu}^\beta(\VEC q)|\tilde 0} \quad ,
\end{split}
\end{equation}
where the sum runs over all possible excited-states of wavevector $\VEC k$ and mode index $r$, and $\ket{\tilde 0}$ is the ground-state of the magnetic system.
We now outline our description of these states, and the detailed derivations are given in Appendix~\ref{sec:adiabatic_approach_of_spin_waves_for_noncolinear_systems}.

We take the generalized Heisenberg Hamiltonian to describe the magnetic system:
\begin{equation}
\label{eq:generalized_hamiltonian}
\begin{split}
  \HH = & -\frac{1}{2} \sum_{ij} \VEC{S}_i^\dagger \VEC J_{ij} \VEC S_j - \sum_i \VEC B_i \cdot \VEC S_i \quad , \\
  \VEC J_{ij}= &
  \begin{pmatrix}
   J_{ij}^x & D_{ij}^z  &-D_{ij}^y \\
  -D_{ij}^z & J_{ij}^y  & D_{ij}^x \\
   D_{ij}^y & -D_{ij}^x & J_{ij}^z+2K^z\delta_{ij} \\
  \end{pmatrix} \quad , \\
  \VEC B_i= &
  \begin{pmatrix}
  B_i^x  & B_i^y  & B_i^z  \\
  \end{pmatrix} \quad .
\end{split}
\end{equation}
$J$ is the isotropic magnetic exchange coupling, which favors collinear alignment for each pair of atomic spins.
$D$ is the antisymmetric Dzyaloshinskii-Moriya interaction, originating from the spin-orbit interaction, that favors a perpendicular alignment.
$B$ is the external magnetic field and $K$ is the uniaxial anisotropy along $z$ (i.e. normal to the lattice plane).
The sum in $i$ and $j$ runs over all magnetic sites of the sample.
The position of each magnetic site can be decomposed by $\VEC{R}_i = \VEC{R}_m + \VEC{R}_\mu$, where $\VEC{R}_m$ and $\VEC{R}_\mu$ are a primitive and a basis vectors, respectively.

The eigenstates of the generalized quantum Heisenberg Hamiltonian are only known for a few special cases.
Thus we have to make some approximations to be able to describe the inelastic scattering from an arbitrary magnetic system.
First we find the ground-state of the classical Hamiltonian, e.g.~by numerical means.
The spin operators $\VEC{S}_\mu$ are given in the global spin frame of reference, with the $z$-axis being normal to the lattice plane.
Then, for every basis atom in the unit cell, we define $\VEC{S}'_\mu$ by a transformation to a local spin frame of reference, where the $z$-axis is given by the classical ground-state spin orientation.
This transformation is represented by a rotation matrix $\VEC{O}_\mu$: $\VEC{S}_\mu=\VEC O_\mu\VEC{S}'_\mu$.

To access the excitation spectrum, we linearize the Holstein-Primakoff representation~\cite{holstein_field_1940} of the quantum spin operators (in the local spin frame of reference): ${S'}^x_\mu-\MR{i}{S'}^y_\mu=\sqrt{2S_\mu}a^\dagger_\mu$, ${S'}^x_\mu+\MR{i}{S'}^y_\mu=\sqrt{2S_\mu}a_\mu$ and $S'^z_\mu = S_\mu - a^\dagger_\mu a_\mu$.
We then truncate the corresponding spin Hamiltonian, keeping only terms up to second order in the Holstein-Primakoff bosons.
The zeroth-order contribution gives the classical ground-state energy, the terms linear in the boson operators vanish when the classical ground-state is used to define the local spin frames, and the quadratic terms describe the spin-excitations.
The lattice Fourier transformation is given by $ a_\mu({\VEC k}) = \frac{1}{\sqrt{N}} \sum_m e^{-\MR{i}\VEC{k} \cdot \VEC{R}_{m}} a_{m\mu} $ and $N$ is the number of unit cells.
Thus, we are left with a Hamiltonian of the form 
\begin{equation}
  \HH_2 =  - \frac{1}{2} \sum_{\VEC k}\sum_{\mu\nu} \VEC a^\dagger_\mu({\VEC k}) \VEC H_{\mu\nu}({\VEC k}) \VEC a_\nu({\VEC k}) \quad , 
\end{equation}
where $\VEC{a}_\mu({\VEC k}) = \begin{pmatrix} a_\mu({\VEC k}) \\ a_\mu^\dagger({-\VEC k}) \\ \end{pmatrix} $, and so $\VEC{H}(\VEC{k})$ is a $2n \times 2n$ matrix with $n$ being the number of atoms in the unit cell.
$\VEC{H}(\VEC{k})$ is generally not block-diagonal for non-collinear systems, so the quadratic part of the Hamiltonian contains `anomalous' terms (in analogy with the theory of superconductivity).
These are eliminated via a Bogoliubov transformation, which diagonalizes $\HH_2$ by introducing a new set of boson operators such that
\begin{equation}
    b^\dagger_r(\VEC{k}) \ket{\tilde0}  = \ket{\VEC{k}r} , \quad
    b_r(\VEC{k}) \ket{\tilde0} = 0 , \quad 
    \braket{\VEC{k}r |\tilde0} = 0 ,
\end{equation}
and
\begin{equation}
  \HH_2 \ket{\VEC{k}r} = \omega_r(\VEC{k}) \ket{\VEC{k}r} \quad.
\end{equation}
The new and old creation and annihilation operators are related by
\begin{equation} 
  a_\mu^\alpha(\VEC k) = \sum_{\beta,r} \mathcal{R}_{\mu r}^{\alpha \beta}(\VEC k) b_{r}^\beta(\VEC k) \quad ,
\end{equation}
The basis transformation matrix $\RR^{\alpha\beta}$ is given by the eigenvectors of the dynamical matrix $\VEC{D}=\VEC{g} \HH_2$, with $\VEC g$ being a diagonal matrix containing $-1$ on its first half and 1 on the second, see Appendix.~\ref{sub:bogoliubov_transformation}.
This development allows us to determine the action of the spin operators on the ground and excited-states of the system, with which, and after some algebra, allows to evaluate Eq.~\ref{eq:density_tensor22} and obtain
\begin{equation}
\begin{split}
     \NN_{\mu\nu}^{\alpha\beta}(\VEC{q}, \omega) =  2\sqrt{ S_\mu S_\nu} \sum_{r} \delta\big(\omega-\omega_{r}(\VEC{q})\big)&  \times \\
       \left[
         O_\mu^{\alpha +}  (\RR_{\mu r}^{++}(\VEC{q}))^*   +
         O_\mu^{\alpha -}  (\RR_{\mu r}^{-+}(\VEC{q}))^*
       \right] & \times \\
      \left[
         O_\nu^{\beta +}  \RR_{\nu r}^{-+}( \VEC{q})   +
         O_\nu^{\beta -}  \RR_{\nu r}^{++}( \VEC{q})  
       \right] & \quad .
\end{split}
\end{equation}
We have then written the spin-spin tensor in terms of the eigenvectors and eigenvalues of the magnetic system.
It is the possibility of accessing different elements of this tensor with a spin-analyzer that provides unique information about the spin-excitations of complex non-collinear magnets, as will be demonstrated in the following.

%***********************************************************************
\textbf{Results}.
We illustrate the significance of this general result with the spin model of Ref.~\onlinecite{Roldan2016}, which was used to describe a magnetic skyrmion lattice on a hexagonal monolayer.
The lattice constant is taken as the unit of length ($a = 1$).
The model consist of only nearest-neighbor interactions, with Dzyaloshinskii-Moriya vectors orthogonal to both the bond direction and the normal to the monolayer plane, $\hat{\VEC{n}}_{ij} = \hat{\VEC{z}} \times \hat{\VEC{R}}_{ij}$.
$B$ is the external magnetic field and $K$ is the uniaxial anisotropy.
The atomic spin is set to $S = 1$ and $J$ is taken as the unit of energy, defining the remaining model parameters as $D = J$, $B = 0.36\,J$ and $K = 0.25\,J$.
We now apply our formalism to three different magnetic states.

\emph{Ferromagnet.} With $D = 0$, the ground state of the spin model is ferromagnetic and its total spin is maximal.
With the polarization of the beam parallel to the spin of the sample, we find only one active inelastic scattering channel, the down-up ($m = -1$).
This is in agreement with the conventional wisdom that it takes a spin-flip process to create a spin-wave, which is the picture familiar from (SP)EELS experiments~\cite{plihal_spin_1999,Vollmer2003,ibach_high_2014}, also found in our previous work~\cite{dos_santos_first-principles_2017}.
For this model, the spectrum features a single and continuous spin-wave branch (Fig.~\ref{fig:dispersionsFMz} in the Appendix~\ref{sec:adiabatic_approach_of_spin_waves_for_noncolinear_systems}).

\begin{figure*}[th!]
  \centering
  \includegraphics[width=1.0\textwidth,trim={0 5 0 0},clip=true]{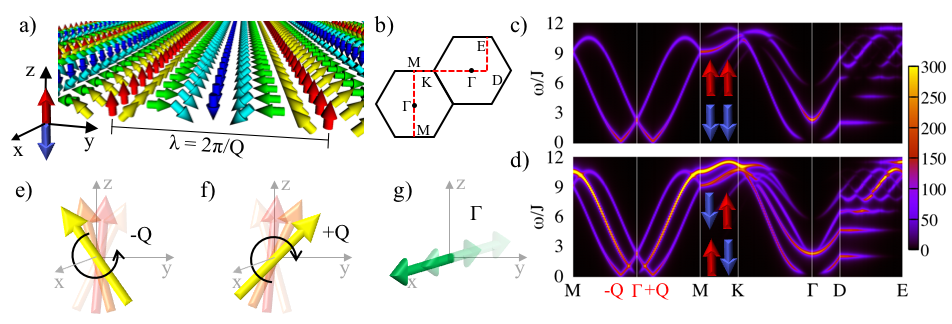}
  \caption{\label{fig:dispersionsSP}
  \textbf{Spin-waves for a spin-spiral structure}, beam polarization along z.
  (a) Spin-spiral ground state and crystallographic axes. The red and blue arrows correspond to the two considered spin polarizations of the electron beam.
  (b) Path in reciprocal space being considered for the calculations of the SREELS spectra.
  These are shown in (c) for the spin-conserving channels, and in (d) for the spin-flip channels.
  The arrow pairs indicate the initial and final electron spin polarization for each channel.
  (e-g) Sketch of the low-frequency motion of the net atomic spin for the three spin-wave modes with minima in $-$Q, $+$Q and $\Gamma$, respectively.
  See also videos 1 to 3 in the Supplementary Materials~\cite{SupplementaryMaterials}.
  }
\end{figure*}

\emph{Spin-spiral.} 
Keeping now only $J$ and $D$ in the spin model, the ground state becomes a spin-spiral.
We considered a cycloidal spin-spiral of wavevector $\VEC Q = Q\,\hat{\VEC{y}}$.
Its energy is minimized by $Q = \alpha/d$, where $\alpha=\arctan(\sqrt{3}D/2J)$ and $d=a\sqrt{3}/2$ is the distance between rows of parallel spins, see Appendix~\ref{apx:ground_state_spiral}.
For convenience, we set $D=2 J/\sqrt{3}$ leading to a spin-spiral wavelength $\lambda = 8d$, as in Fig.~\ref{fig:dispersionsSP}(a).
This magnetic state has zero net magnetization.

Fig.~\ref{fig:dispersionsSP}(c-d) shows the spin-resolved inelastic electron scattering spectra calculated from Eq.~\eqref{eq:transrate} on the path of Fig.~\ref{fig:dispersionsSP}(b).
We considered the electron beam polarization along $z$ --- up and down are defined with respect to this axis.
The spin-conserving channels ($m=0$) always present the same response, because they measure excitations that have zero net angular momentum and, therefore, are insensitive to the spin of the probing electrons.
Here, the spin-flip channels are equivalent because of the symmetry of the magnetic structure with respect to $z$.
Three modes are clearly observed in the spin-flip channels, Fig.~\ref{fig:dispersionsSP}(d), as sharp and well-defined dispersing features through the M--$\Gamma$--M path.
They have energy minima in $-$Q, $\Gamma$ and $+$Q, which we will use to label them.
These modes are the three universal helimagnon modes~\cite{janoschek_helimagnon_2010}, in contrast to the single Goldstone mode in ferromagnets.
For low frequency, the $-$Q and $+$Q are excitations that yield a net atomic spin rotating counter-clockwise and clockwise, respectively, in the $z-y$ plane, see Fig.~\ref{fig:dispersionsSP}(e-f).
For the $\Gamma$-mode, however, the total atomic spin does not rotate but oscillates linearly along the $x$-axis, as in Fig.~\ref{fig:dispersionsSP}(g).
This shows that non-collinear magnetic structures can host zero net angular momentum spin-waves and that they can be observed by SREELS.
Note yet the highly anisotropic dispersion-relation around the $\Gamma$-point.
It is linear or quadratic for spin-waves propagating parallel or transversal to $\VEC Q$, respectively, as seen in Fig.~\ref{fig:dispersionsSP}(d), paths M--$\Gamma$--M and K--$\Gamma$--D~\cite{belitz_theory_2006,maleyev_cubic_2006}.
Furthermore, Fig.~\ref{fig:dispersionsSP}(c-d, path D-E) shows the formation of one-dimensional spin-waves, as indicated by the dispersionless bands~\cite{janoschek_helimagnon_2010,fischer_weak_2004}.

\begin{figure}[b!]
  \centering
  \includegraphics[width=0.5\textwidth,trim={0em 1.5em 0 0.2em},clip=true]{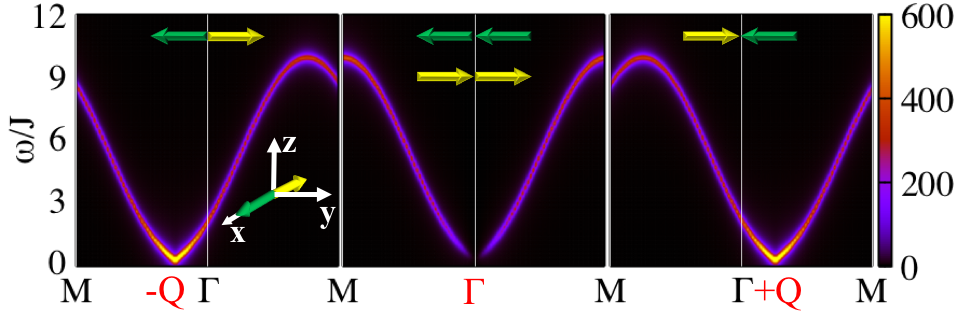}
  \caption{\label{fig:spiral_selection}
  SREELS spectra for spin-waves in a spin-spiral as in Fig.~\ref{fig:dispersionsSP}.
  Here, the beam polarization is along $x$, which is aligned with the precession axis of the spin-waves.
  Thus, each scattering channel probes a single spin-wave mode.
  }
\end{figure}

\begin{figure*}[t!]
  \centering
  \includegraphics[width=1.0\textwidth,trim={7 4 5 4},clip=true]{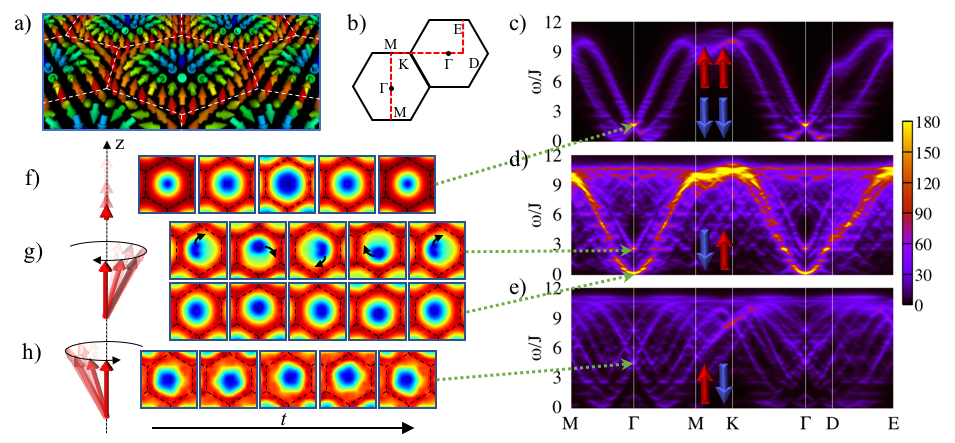}
  \caption{\label{fig:dispersionsSK}
  \textbf{Spin-waves in a skyrmion lattice.}
  (a) Shows the ground state spin structure of the system. The colors represent the z-component of the spins.
  (b) Depicts the path on which all four SREELS spectra were calculated, (c-e).
  (f--h) snapshots of the $z$-component of the local atomic spins over time (as color maps), depicting the spin-wave motion at the hotspots of the spectra. Same color scale as in (a).
  (f) corresponds to a breathing mode that is measured in the non-spin-flip channels.
  (g) and (h) are clockwise and counter-clockwise rotational modes observed in the down-up and up-down channels, respectively.
  See also videos 4 to 8 in the Supplementary Materials~\cite{SupplementaryMaterials}.
  }
\end{figure*}

The dynamics of the spin-wave modes depicted in Fig.~\ref{fig:dispersionsSP}(e-g) indicates the $x$-axis as the natural quantization axis.
It defines left and right spin projections.
An incident electron with up or down polarization corresponds to a superposition of left and right spinors with respect to the $x$-axis.
The $-$Q ($+$Q) mode can be excited by an electron with left (right) polarization, which then undergoes a spin-flip and goes out with right (left) polarization.
Therefore, $-$Q and $+$Q are seen by the spin detector as a superposition of the up and down polarizations, and this makes them to be detected in all channels.
Due to quantum interference the $\Gamma$-mode disappears from the non-spin-flip channels, and it is intensified in the spin-flip ones, see Fig.~\ref{fig:dispersionsSP}(c-d).
Now, if we rotate the polarization of the electron beam to be aligned with the $x$-axis, each mode will appear in a distinct scattering channel, as demonstrated in Fig.~\ref{fig:spiral_selection}.
Also, overall the intensities are higher now with the polarization axis along the spin-wave precession axis.
In practice, controlling the polarization direction of the beam and of the spin detector, which indeed are independent, allows SREELS to select or render undetected certain spin-wave modes.

\emph{Skyrmion lattice}.
An increasing external magnetic field is responsible for deforming the spin-spiral phase into a conical state, then into the skyrmion lattice~\cite{dupe_tailoring_2014}.
We concentrate on the skyrmion lattice phase shown in Fig.~\ref{fig:dispersionsSK}(a), which was obtained via a numerical energy minimization including all the model parameters.
The polarization of the electron beam is again along $z$.
Fig.~\ref{fig:dispersionsSK}(c-e) shows the SREELS spectra on the path displayed in Fig.~\ref{fig:dispersionsSK}(b).
Fig.~\ref{fig:dispersionsSK}(c) demonstrates that the spin-wave spectrum of a skyrmion lattice inherits the two-mode structure found for the spin-spiral, see Fig.~\ref{fig:dispersionsSP}(c), although both branches are now much broader.
Contrary to the usual spin-wave broadening due to coupling to phonons or electrons~\cite{dai_magnon_2000,muniz_microscopic_2003,costa_theory_2004}, here it originates in the non-collinearity of the magnetization.
Note that the down-up spectrum in Fig.~\ref{fig:dispersionsSK}(d) has overall a higher intensity than the up-down one in Fig.~\ref{fig:dispersionsSK}(e), due to the upward total atomic spin of the system.
Still in Fig.~\ref{fig:dispersionsSK}(d), around $\Gamma$ we observe that the gapless feature has a quadratic dispersion, while the one with minimum at $\omega / J \sim 3 $ disperses linearly.

Fig.~\ref{fig:dispersionsSK}(f-h) depicts the time evolution of the spin-wave modes responsible for the high intensity spots at the $\Gamma$-point in the various channels.
The color maps represent the $z$-component of the local atomic spins, and the arrows illustrate the total atomic spin.
The hotspot in the non-spin-flip channels, Fig.~\ref{fig:dispersionsSK}(c), is due to a breathing mode, where the skyrmion core shrinks and enlarges periodically.
It has zero net angular momentum, as seen by the dynamics of the total atomic spin in Fig.~\ref{fig:dispersionsSK}(f).
Two rotational modes identified in the down-up channel near $\omega / J \sim 3 $ and at zero are clockwise, and the dynamics of their total atomic spin indicates that they possess downward angular moments, Fig.~\ref{fig:dispersionsSK}(g).
A counter-clockwise rotational mode is responsible for the faint hotspot in the up-down channel, Fig.~\ref{fig:dispersionsSK}(h), therefore, with upward angular momentum.
This explains their appearance in their respective scattering channels.

%***********************************************************************
\textbf{Discussion and conclusions}.
We showed that inelastic electron scattering can reveal various spin-wave phenomena in non-collinear magnets throughout the reciprocal space.
We demonstrated that it can measure anisotropies in the dispersion relation, and the localization of spin-waves along certain directions that yields to desired spin-wave channeling for spintronics~\cite{belitz_theory_2006,maleyev_cubic_2006,janoschek_helimagnon_2010,fischer_weak_2004}.
Furthermore, we discovered that the spin-analysis of the scattered electrons gives access to novel properties of the spin-waves in non-collinear substrates, such as zero net angular momentum modes.
Also, manipulating the polarization of the electron beam allows to select and filter spin-wave modes.

The realization of the SREELS may be applied to fingerprint magnetic phases from their unique signatures on the spin-wave spectra.
It could, for example, help to distinguish between a skyrmion tube and a magnetic bobber lattice in thin films~\cite{zheng_experimental_2017}.
These phases may have similar magnetic profiles at the very surface, but they differ deeper inside the film, which impacts on the spin-waves.
Also, our theoretical approach can be straightforward applied for material specific predictions, if magnetic interaction parameters obtained from first-principles calculations are supplied.

The presence of spin-orbit coupling in the magnetic sample leads to an additional source of spin-dependent scattering, besides the exchange scattering mechanism that lets us probe spin-excitations.
An incoming electron can scatter on the spin-orbit potential and have its spin flipped.
Subsequently, it can then further create or annihilate spin-waves, which contributes to the inelastic signal.
For magnetic transition metals, spin-orbit coupling is much weaker than the exchange scattering, which is why such scattering processes can be ignored.
On heavy metals, spin-orbit coupling becomes important, e.g., allowing a full spin characterization of flying electrons by surface skew-scattering~\cite{schaefer_vectorial_2017}.

This could be used as a spin-filter in SREELS, but it might also require a high intensity electron beam, to compensate for the low efficiency of the spin detection.

A more efficient and intuitive way for spin-filtering could be a Stern-Gerlach apparatus for electrons.
However, the feasibility of such an experiment has been discussed since the earliest years of quantum mechanics, and it is still a matter of debate~\cite{batelaan_stern-gerlach_1997,rutherford_comment_1998,garraway_observing_1999,garraway_does_2002,larson_transient_2004,karimi_spin--orbital_2012}.
We hope that our work will encourage investigations on such a spin-splitter by providing an important application, having in mind that similar challenges have been overcome for neutron scaterring experiment~\cite{moon_polarization_1969}.
Despite the enrichment that the spin analysis brings to the discussion, spin-waves in non-collinear systems can be measured with the existing (SP)EELS setups.
Their spectra would consist of combinations of the different scattering channels we have described for SREELS.

%***********************************************************************
\textbf{Acknowledgments.} We thank J.~Azpiroz, J.~Chico and H.~Ibach for the critical reading of our manuscript.
This work is supported by the Brazilian funding agency CAPES under Project No. 13703/13-7 and the European Research Council (ERC) under the European Union's Horizon 2020 research and innovation programme (ERC-consolidator Grant No. 681405-DYNASORE).

\appendix

% \graphicspath{ {Supplementary_Materials/figures/} }

\tableofcontents

\section{Inelastic electron scattering theory} % (fold)
\label{Apx:inelastic_scattering}
Here we present the derivation of the transition rate for inelastic electron scattering from spin waves of magnetic systems, Eq.~(1) of the main text.
The complete theory of electron diffraction from a surface is highly involved, due to the strong interaction of the beam electrons with those of the sample.
However, as our interest is in the inelastic signal from magnetic origin, we shall simplify the problem by treating the surface as a lattice of atomic spins in their ground state, with a local spin exchange interaction describing the coupling to the beam electrons.
In the next section, we will discuss the particularity of applying this theory for non-collinear magnets.

\subsection{General framework}
The Hamiltonian of the problem has the following parts:
\begin{equation}
\begin{split}
  &\MC{H}_{\MR{e}} = \frac{\VEC{p}^2}{2m_{\MR{e}}} \quad, \\
  &\MC{H}_{\MR{m}} = -\frac{1}{2} \sum_{mn}\sum_{\alpha\beta} S_m^\alpha J_{mn}^{\alpha\beta} S_n^\beta - \sum_{n}\sum_{\alpha} B_n^\alpha S_n^\alpha \quad, \\
  &\MC{H}_{\MR{em}} = \sum_n\sum_\alpha U_n\,\delta(\VEC{r} - \VEC{R}_n)\,\sigma^\alpha S_n^\alpha \quad .
\end{split}
\end{equation}
The electron beam is described by the free-electron Hamiltonian $\MC{H}_{\MR{e}}$, with $\VEC{p}$ the linear momentum operator and $m_{\MR{e}}$ the electron mass.
The magnetic lattice is described by $\MC{H}_{\MR{m}}$, with $S_n^\alpha$ being the $\alpha$-component of the atomic spin operator for site $n$, $J_{mn}^{\alpha\beta}$ the elements of the tensor describing the pairwise interactions between sites $m$ and $n$, and $B_n^\alpha$ the $\alpha$-component of the magnetic field acting on site $n$.
The coupling between the atomic spins and the spin of the beam electrons is described by $\MC{H}_{\MR{em}}$, with $U_n$ the interaction strength, $\VEC{r}$ the position operator for the electrons, $\VEC{R}_n$ the position vector for site $n$, and $\sigma^\alpha$ the Pauli matrix for the $\alpha$-component of the electron spin.

Next, we assume that the beam electrons and the magnetic sample are decoupled for times $t < 0$.
Then we can specify the initial state of the electron beam as consisting of a plane-wave with well-defined energy $E_i$, wavevector $\VEC{k}_i$ and spin $s_i$,
\begin{equation}\label{eq:spinpol}
\begin{split}
  &\braket{\VEC{r}|\VEC{k}_i s_i} = e^{\iu \VEC{k}_i\cdot\VEC{r}} \ket{s_i}  \quad, \quad
  E_i = \frac{\VEC{k}_i^2}{2m} \quad,\\
  &\ket{s_i} \! \bra{s_i} = \frac{1}{2}\,(\sigma^0 + \VEC{n}_i \cdot \boldsymbol{\upsigma}) \quad .
\end{split}
\end{equation}
Henceforth $\hbar =1$.
The spinor $\ket{s_i}$ defines the spin polarization of the electron to be along the direction $\VEC{n}_i$.
The eigenstates of the spin model are assumed to be known,
\begin{equation}
  \MC{H}_{\MR{m}} \ket{\lambda} = E_\lambda \ket{\lambda} \quad,\qquad E_0 \leq E_\lambda \quad,
\end{equation}
and the magnetic sample is in its ground state $\ket{0}$, with energy $E_0$.
The state of the combined system at $t = 0$ is then the tensor product of the two initial states
\begin{equation}
  \ket{i} \equiv \ket{\VEC{k}_i s_i 0} = \ket{\VEC{k}_i s_i} \otimes \ket{0} \quad .
\end{equation}
This state evolves in time under the action of the complete Hamiltonian $\MC{H} = \MC{H}_{\MR{e}} + \MC{H}_{\MR{m}} + \MC{H}_{\MR{em}}$, according to the Schr\"odinger equation,
\begin{equation}
  \iu\frac{\ud}{\ud t}\ket{\Psi(t)} = \MC{H}\ket{\Psi(t)} \quad,\qquad \ket{\Psi(0)} = \ket{i} \quad .
\end{equation}
We introduce the time evolution operator, that connects the state at a later time $t$ to the initial state, in the form
\begin{equation}
\begin{split}
 & \ket{\Psi(t)} = e^{-\iu \MC{H}_0 t}\,\MC{U}(t)\ket{i} \quad\Longrightarrow\quad \\
 & \MC{U}(t) = 1 - \iu\int_0^t\!\!\ud t_1\;\MC{H}_{\MR{em}}(t_1)\;\MC{U}(t_1) \quad, \\
 & \MC{H}_{\MR{em}}(t) = e^{\iu \MC{H}_0 t}\,\MC{H}_{\MR{em}}\,e^{-\iu \MC{H}_0 t} \quad .
\end{split}
\end{equation}
This integral equation follows directly from the Schr\"odinger equation.
The total Hamiltonian is split as $\MC{H} = \MC{H}_0 + \MC{H}_{\MR{em}}$, with $\MC{H}_0 = \MC{H}_{\MR{e}} + \MC{H}_{\MR{m}}$.
Iterating the integral equation, we find
\begin{equation}
\begin{split}
  \MC{U}(t) =& 1 - \iu\int_0^t\!\!\ud t_1\;\MC{H}_{\MR{em}}(t_1)
  +   \\
  &(-\iu)^2\!\int_0^t\!\!\ud t_1 \int_0^{t_1}\!\!\!\!\ud t_2\; \MC{H}_{\MR{em}}(t_1)\;\MC{H}_{\MR{em}}(t_2) + \ldots \\
  =& 1 + \MC{U}_1(t) + \MC{U}_2(t) + \ldots
\end{split}
\end{equation}
This expansion corresponds to performing time-dependent perturbation theory in $\MC{H}_{\MR{em}}$.
%The operators $\MC{U}(t)$ and $\MC{U}_0(t)$ are unitary, but the other $\MC{U}_n(t)$ are not.
%To obtain the expansion of $\MC{U}^\dagger(t) = \MC{U}(-t)$, replace $t \rightarrow -t$ in the corresponding expressions.

The probability of finding the system at a later time in some final state $\ket{f} = \ket{\VEC{k}_{\!f} s_{\!f} \lambda}$ is
\begin{equation}
\begin{split}
  &P(i\!\rightarrow\!f,t) = \big|\!\braket{f|\Psi(t)}\!\big|^2
  = \braket{i | \,\MC{U}^\dagger(t) | f}\!\braket{f | \,\MC{U}(t) | i} \\
  &\approx \big|\!\braket{f | i}\!\big|^2 \quad \big(= P_0(i\!\rightarrow\!f,t)\big) \\
  &+ \left( \braket{i | f} \! \braket{f | \,\MC{U}_1(t) | i} + \braket{i | \,\MC{U}_1^\dagger(t) | f} \! \braket{f |i} \right) \quad \big(= P_1(i\!\rightarrow\!f,t)\big) \\
  &+ \left( \braket{i | \,\MC{U}_1^\dagger(t) | f} \! \braket{f | \,\MC{U}_1(t) | i} 
  + \braket{i | f} \! \braket{f | \,\MC{U}_2(t) | i} \right. + \\
  & \qquad \qquad \qquad \left. + \braket{i | \,\MC{U}_2^\dagger(t) | f} \! \braket{f | i} \right)  \quad \big(= P_2(i\!\rightarrow\!f,t)\big) 
  \quad ,
\end{split}
\end{equation}  
up to second order in $\MC{H}_{\MR{em}}$.
Conservation of probability leads to
\begin{equation}
\begin{split}
  &\sum_f P(i\!\rightarrow\!f,t) = 1  , \quad  
   \sum_f P_0(i\!\rightarrow\!f,t) = 1 \\
  & \Longrightarrow \quad
   \sum_f P_n(i\!\rightarrow\!f,t) = 0 , \quad n > 0 \quad .
\end{split}
\end{equation}

The transition amplitudes are ($E_b - E_a \equiv E_{ba}$)
\begin{equation}
\begin{split}
  \braket{f | \,\MC{U}_1(t) | i} &= -\iu\int_0^t\!\!\ud t_1 \braket{f | e^{\iu \MC{H}_0 t_1}\,\MC{H}_{\MR{em}}\,e^{-\iu \MC{H}_0 t_1} | i} \\
  &= \frac{1 - e^{\iu E_{\!fi}t}}{E_{\!fi}}\,\braket{f | \MC{H}_{\MR{em}} | i} \quad ,
\end{split}
\end{equation}
\begin{equation}
\begin{split}
  \braket{f | \,\MC{U}_2(t) | i} &= -\sum_v \int_0^t\!\!\ud t_1 \int_0^{t_1}\!\!\!\!\ud t_2 \braket{f | e^{\iu \MC{H}_0 t_1}\,\MC{H}_{\MR{em}} | v}\! \times \\
  & \quad \qquad \qquad \braket{v | e^{-\iu \MC{H}_0 (t_1-t_2)}\,\MC{H}_{\MR{em}}\,e^{-\iu \MC{H}_0 t_2} | i} \\
  &= \sum_v \left(\frac{e^{\iu E_{\!fi} t} - 1}{E_{fi}\,E_{vi}} - \frac{e^{\iu E_{\!fv} t} - 1}{E_{fv}\,E_{vi}} \right)
  \braket{f | \MC{H}_{\MR{em}} | v}\! \times \\
  & \quad \qquad \qquad  \quad \qquad \qquad   \qquad \qquad \braket{v | \MC{H}_{\MR{em}} | i} \quad .
\end{split}
\end{equation}
A complete set of (virtual) states was introduced for the second-order amplitude.

The zeroth-order contribution to the transition probability is
\begin{equation}
  P_0(i\!\rightarrow\!f,t) = \left|\braket{f | i}\right|^2 \quad .
\end{equation}
The final state must have a finite overlap with the initial state for a non-vanishing result.
As $\ket{f} = \ket{\VEC{k}_{\!f} s_{\!f} \lambda}$, this requires $\VEC{k}_{\!f} = \VEC{k}_i$ and $\lambda = 0$.
The spinors give, see Eq.~\eqref{eq:spinpol},
\begin{align}
  P_0(i\!\rightarrow\!f,t) = & \left|\braket{s_{\!f} | s_i}\right|^2 = \frac{1}{4}\,\Tr\,(\sigma^0 + \VEC{n}_i \cdot \boldsymbol{\upsigma})\,(\sigma^0 + \VEC{n}_{\!f} \cdot \boldsymbol{\upsigma}) \nonumber\\
   = & \frac{1}{2}\,(1 + \VEC{n}_i \cdot \VEC{n}_{\!f}) \quad .
\end{align}
Measuring the spin component of the outgoing electron with a spin detector which is not aligned with the polarization of the incident electron beam then leads to a cosine dependence on the angle between them.

The first-order contribution to the transition probability is
\begin{equation}
\begin{split}
  P_1(i\!\rightarrow\!f,t) = & \frac{1 - e^{\iu E_{\!fi} t}}{E_{\!fi}} \braket{i|f} \! \braket{f | \MC{H}_{\MR{em}} | i} \\
  & + \frac{1 - e^{-\iu E_{\!fi} t}}{E_{\!fi}}\,\braket{i | \MC{H}_{\MR{em}} | f} \! \braket{f|i} \quad ,
\end{split}
\end{equation}
and the respective scattering rate is (recall that $\braket{f | i}$ must be finite, so $E_{\!fi} \rightarrow 0$)
\begin{equation}
\begin{split}
  \Gamma_1(i\!\rightarrow\!f,t) &= \frac{\ud P_1}{\ud t}(i\!\rightarrow\!f,t) \\
    &= -\iu\,\big(\!\braket{s_i | s_{\!f}} \! \braket{\VEC{k}_i s_{\!f} 0 | \MC{H}_{\MR{em}} | \VEC{k}_i s_i 0} \\
    & \qquad \qquad - \braket{\VEC{k}_i s_i 0 | \MC{H}_{\MR{em}} | \VEC{k}_i s_{\!f} 0} \! \braket{s_{\!f} | s_i}\!\big) \\
    &= \sum_n U_n \braket{0 | \VEC{S}_n | 0} \cdot (\VEC{n}_i \times \VEC{n}_{\!f}) \quad .
\end{split}
\end{equation}
Its detection requires a crossed setup: the polarization of the outgoing electron must be measured along a direction perpendicular to the polarization of the incident beam, yielding information about the component of the magnetization of the sample perpendicular to those two axes.

The second-order contribution is the most interesting one, as it describes inelastic scattering.
The first contribution to the transition probability is
\begin{equation}
  P_{2,1}(i\!\rightarrow\!f,t) = 2\,\frac{1 - \cos(E_{\!fi} t)}{(E_{\!fi})^2} \left|\braket{f | \MC{H}_{\MR{em}} |i}\right|^2 \quad,
\end{equation}
with the scattering rate
\begin{equation}
\begin{split}
  \Gamma_{2,1}(i\!\rightarrow\!f,t) & = \frac{\ud P_{2,1}}{\ud t}(i\!\rightarrow\!f,t)
  = 2\,\frac{\sin(E_{\!fi} t)}{E_{\!fi}} \left|\braket{f | \MC{H}_{\MR{em}} |i}\right|^2 \\
 & \underset{t \rightarrow \infty}{=} 2\pi\,\delta(E_{\!fi}) \left|\braket{f | \MC{H}_{\MR{em}} |i}\right|^2 \quad .
\end{split}
\end{equation}
This is the familiar Fermi's Golden Rule.
The delta function imposes energy conservation:
\begin{equation}
  0 = E_{\!fi} = E_\lambda + \frac{\VEC{k}_{\!f}^2}{2m} - E_0 - \frac{\VEC{k}_i^2}{2m} = E_\lambda - E_0 - \omega \quad,
\end{equation}
with $\omega = E_\lambda - E_0$ the energy transferred from the electron beam to the magnetic sample.
Likewise, we can define $\VEC{q} = \VEC{k}_i - \VEC{k}_{\!f}$ as the momentum transferred to the magnetic sample.

There is another contribution in second order,
\begin{equation}
\begin{split}
  P_{2,2}(i\!\rightarrow\!f,t) &= \sum_v \left(\frac{e^{\iu E_{\!fi} t} - 1}{E_{fi}\,E_{vi}} - \frac{e^{\iu E_{\!fv} t} - 1}{E_{fv}\,E_{vi}} \right)
  \braket{i | f} \times \\
  & \qquad \qquad \qquad \! \braket{f | \MC{H}_{\MR{em}} | v}\!\braket{v | \MC{H}_{\MR{em}} | i} + \\
  &+ \sum_v \left(\frac{e^{-\iu E_{\!fi} t} - 1}{E_{fi}\,E_{vi}} - \frac{e^{-\iu E_{\!fv} t} - 1}{E_{fv}\,E_{vi}} \right) \times \\
  & \qquad \qquad \qquad  \braket{i | \MC{H}_{\MR{em}} | v}\!\braket{v | \MC{H}_{\MR{em}} | f} \! \braket{f | i} \quad .
\end{split}
\end{equation}
Due to the presence of the overlap $\braket{f | i}$, it contributes only to $\omega = 0$ and $\VEC{q} = \VEC{0}$.
As we are interested in inelastic scattering, we will not analyze this term further.

\subsection{Inelastic scattering rate} % (fold)
\label{sub:scattering_rate}

From the analysis in the previous section, we can define the inelastic scattering rate as expected from Fermi's Golden Rule:
\begin{equation}
  \Gamma_{if}(\VEC{q},\omega) = 2\pi \sum_{\lambda \neq 0} \delta(E_\lambda - E_0 - \omega)\,\left|\braket{\VEC{k}_{\!f} s_{\!f} \lambda | \MC{H}_{\MR{em}} |\VEC{k}_i s_i 0}\right|^2 \quad ,
\end{equation}
with $\omega$ and $\VEC{q}$ the energy and momentum transferred from the electron beam to the magnetic sample.

We assume that the ground state of the magnetic sample is commensurate with the atomic lattice, and for simplicity consider a single monolayer.
Then we can separate the position vector of every magnetic atom as $\VEC{R}_{n\nu} = \VEC{R}_n + \VEC{R}_\nu$, letting $\VEC{R}_n$ label the origin of the $n$-th magnetic unit cell, and $\VEC{R}_\nu$ the basis vector inside the magnetic unit cell.
The coupling Hamiltonian is assumed to have the translational symmetry of the magnetic unit cell, so
\begin{equation}
  \MC{H}_{\MR{em}} = \sum_{n\nu} U_\nu\,\delta(\VEC{r} - \VEC{R}_{n\nu})\,\bm{\upsigma}\cdot\VEC{S}_{n\nu} \quad .
\end{equation}
If the magnetic atoms are chemically distinct, their coupling strength might be atom-dependent, hence $U_\nu$.
The matrix elements are then
\begin{equation}
\begin{split}
  \braket{\VEC{k}_{\!f} s_{\!f} \lambda | \MC{H}_{\MR{em}} | \VEC{k}_i s_i 0}
  = & \sum_\beta \braket{s_{\!f} | \sigma^\beta | s_i} \sum_{\nu} U_\nu\,e^{\iu\VEC{q}\cdot\VEC{R}_\nu} \times \\
    & \qquad \qquad \qquad  \braket{\lambda | \sum_n e^{\iu\VEC{q}\cdot\VEC{R}_n} S_{n\nu}^\beta | 0} \\
  = & \sqrt{N_l}\,\sum_\beta \braket{s_{\!f} | \sigma^\beta | s_i} \sum_{\nu} U_\nu\,e^{\iu\VEC{q}\cdot\VEC{R}_\nu} \times\\
    & \qquad \qquad \qquad  \braket{\lambda | S_\nu^\beta(\VEC{q}) | 0} \quad ,
\end{split}
\end{equation}
\begin{equation}
\begin{split}
  \braket{\VEC{k}_i s_i 0 | \MC{H}_{\MR{em}} | \VEC{k}_{\!f} s_{\!f} \lambda} &= \sum_\alpha \braket{s_i | \sigma^\alpha | s_{\!f}} \sum_{\mu} U_\mu\,e^{-\iu\VEC{q}\cdot\VEC{R}_\mu} \times \\ 
  & \qquad \qquad \qquad  \braket{0 | \sum_n e^{-\iu\VEC{q}\cdot\VEC{R}_m} S_{m\nu}^\alpha | \lambda} \\
  &= \sqrt{N_l}\,\sum_\alpha \braket{s_i | \sigma^\alpha | s_{\!f}} \sum_{\mu} U_\mu\,e^{-\iu\VEC{q}\cdot\VEC{R}_\mu} \times \\
  &  \qquad \qquad \qquad  \braket{0 | S_{\mu}^\alpha(-\VEC{q}) | \lambda} \quad .
\end{split}
\end{equation}
$N_l$ is the number of unit cells under Born-von Karman periodic boundary conditions.
We define the spin-spin correlation tensor as
\begin{equation} \label{eq:spinspin_tensor}
  \MC{N}_{\mu\nu}^{\alpha\beta}(\VEC{q},\omega) = \sum_{\lambda \neq 0} \delta(E_\lambda - E_0 - \omega)
  \braket{0 | S_\mu^\alpha(-\VEC{q}) | \lambda} \! \braket{\lambda | S_\nu^\beta(\VEC{q}) | 0} \quad .
\end{equation}
It has the periodicity of the magnetic lattice, with $\alpha,\beta = x,y,z$ the components of the spin operators, and describes the intrinsic spin excitations of the magnetic sample.

The inelastic scattering rate is then expressed using this tensor as
\begin{equation} \label{eq:scattering_rate}
\begin{split}
  \Gamma_{if}(\VEC{q},\omega) =& 2\pi\,N_l\,N_b \sum_{\alpha\beta} \braket{s_i | \sigma^\alpha | s_{\!f}} \! \braket{s_{\!f} | \sigma^\beta | s_i} \times \\
 & \qquad  \frac{1}{N_b}\sum_{\mu\nu} U_\mu U_\nu\,e^{\iu\VEC{q}\cdot\VEC{R}_{\mu\nu}} \MC{N}_{\mu\nu}^{\alpha\beta}(\VEC{q},\omega) \quad ,
\end{split}
\end{equation}
with $\VEC{R}_{\mu\nu} = \VEC{R}_\nu - \VEC{R}_\mu$.
$N_b$ is the number of basis atoms in each unit cell, so $N_l N_b$ is the total number of magnetic atoms.
The scattering rate combines the information about the intrinsic spin excitations, contained in $\MC{N}_{\mu\nu}^{\alpha\beta}(\VEC{q},\omega)$, with the information about the spin polarization of the incoming and detected electrons (Pauli matrices) and the wave nature of the electrons, leading to interference between different contributions (the Fourier phase factor).

We can find an explicit expression for the dependence on the electron spin polarization:
\begin{equation}
\begin{split}
   \MC{P}_{if}^{\alpha\beta} =& \braket{s_i | \sigma^\alpha | s_{\!f}} \! \braket{s_{\!f} | \sigma^\beta | s_i} \\
   %= \Tr\,\ket{s_i} \! \bra{s_i}  \sigma^\alpha \ket{s_{\!f}} \! \bra{s_{\!f}} \sigma^\beta \\
   =& \frac{1}{4}\,\Tr\,\big(\sigma^0 + \VEC{n}_i\cdot\bm{\upsigma}\big)\,\sigma^\alpha\,\big(\sigma^0 + \VEC{n}_{\!f}\cdot\bm{\upsigma}\big)\,\sigma^\beta \\
   =&\frac{1}{2}\,\Big(\big(1 - \sum_\gamma n_i^\gamma\,n_{\!f}^\gamma\big)\,\delta_{\alpha\beta} + n_i^\alpha\,n_{\!f}^\beta + n_i^\beta\,n_{\!f}^\alpha + \\
   & \qquad \qquad  \qquad+ \iu\sum_\gamma \varepsilon_{\alpha\beta\gamma}\,\big(n_i^\gamma - n_{\!f}^\gamma\big)\Big) \quad .
\end{split}
\end{equation}
Here $\delta_{\alpha\beta}$ is the usual Kronecker delta, and $\epsilon_{\alpha\beta\gamma}$ the Levi-Civita symbol.
To illustrate, consider the spin polarization of the incoming electrons to be $+z$ or $-z$, and the spin polarization of the outgoing electrons also to be measured along $+z$ or $-z$.
The four tensors selecting the spin components of the magnetic sample that can be measured for each case are
\begin{equation}
\begin{split}
  \renewcommand{\arraystretch}{1.2}
  &\begin{array}{c | c c c}
    \MC{P}_{++} & x & y & z \\
    \hline
    x & 0 & 0 & 0 \\
    y & 0 & 0 & 0 \\
    z & 0 & 0 & 1
  \end{array} \quad,\qquad
  \begin{array}{c | c c c}
    \MC{P}_{--} & x & y & z \\
    \hline
    x & 0 & 0 & 0 \\
    y & 0 & 0 & 0 \\
    z & 0 & 0 & 1
  \end{array} \quad,\\
  &\begin{array}{c | c c c}
    \MC{P}_{+-} & x & y & z \\
    \hline
    x & 1 & +\iu & 0 \\
    y & -\iu & 1 & 0 \\
    z & 0 & 0 & 0
  \end{array} \quad,\qquad
  \begin{array}{c | c c c}
    \MC{P}_{-+} & x & y & z \\
    \hline
    x & 1 & -\iu & 0 \\
    y & +\iu & 1 & 0 \\
    z & 0 & 0 & 0
  \end{array} \quad.
\end{split}
\end{equation}
We see that $\MC{P}_{++}$ and $\MC{P}_{--}$ are the same, and connect with $\MC{N}_{\mu\nu}^{zz}(\VEC{q},\omega)$.
$\MC{P}_{+-}$ connects with $\MC{N}_{\mu\nu}^{-+}(\VEC{q},\omega)$, and $\MC{P}_{-+}$ connects with $\MC{N}_{\mu\nu}^{+-}(\VEC{q},\omega)$.

For a ferromagnetic sample with a ground state of total spin along $+z$, only $\MC{N}_{\mu\nu}^{+-}(\VEC{q},\omega)$ is finite.
$\MC{P}_{-+}$ means that the spin polarization of the incoming electron beam is $-z$, antiparallel to the total spin of the sample.
As the outgoing electron is detected with $+z$ spin polarization, the ferromagnetic sample lost $\hbar$ of angular momentum, corresponding to the lowering of the spin associated with the creation of a spin wave.
If $\MC{N}_{\mu\nu}^{-+}(\VEC{q},\omega)$ were finite, then the sample would gain $\hbar$ of angular momentum.
More intriguingly, a finite $\MC{N}_{\mu\nu}^{zz}(\VEC{q},\omega)$ describes spin excitations with no net exchange of angular momentum between electron beam and magnetic sample.

% subsection spin_spin_correlation_tensor_for_non_colinera_magnets (end)

\section{Adiabatic approach of spin waves for noncolinear systems} % (fold)
\label{sec:adiabatic_approach_of_spin_waves_for_noncolinear_systems}

Our goal is to calculate the inelastic scattering rate when an electron beam scatters from spin-waves in a non-collinear magnet.
This rate is given by Eq.~\eqref{eq:scattering_rate}, and therefore we need to evaluate the spin-spin correlation tensor of Eq.~\eqref{eq:spinspin_tensor}.
For that, we need to determine the ground state $\ket{0}$ of the system, and describe its excited states $\ket{\lambda}$.
Also, we need to establish how the spin operators act on these states.
As example cases, we are going to consider two magnetic phases of a monolayer hexagonal crystal: a spin-spiral and a skyrmion lattice.
To describe the magnetic system, we consider the generalized Heisenberg Hamiltonian of Eq.~\ref{eq:generalized_hamiltonian}.
The first step consists of determining the classical ground-state spin-configuration.

\subsection{The classical ground state} \label{sec:ground_state}

The classical ground state of a magnetic system is given by the configuration of the classical spins that has the lowest total energy.
To find such a configuration, we replace the spin operators $\VEC S_i$ by classical vectors in the Hamiltonian of Eq.~\eqref{eq:generalized_hamiltonian}, then we search for the spin alignments with respect to each other and to the fields that is energetic mostly favorable, considering that the magnitude of $\VEC S_i$ is constant.
This search is not a trivial matter in general, and it can be attempted analytically or numerically depending on the complexity of the set of interactions.
We want now to determine the classical ground state of a spin-spiral and a skyrmion lattice for a hexagonal monolayer.

\subsubsection{Spin-spiral}
\label{apx:ground_state_spiral}

Now let us consider a single layer of a hexagonal lattice of primitive vectors $\VEC a_1=a\hat{ \VEC{x}}$ and $\VEC a_2=a(\hat{ \VEC{x}}/2+\sqrt{3}\hat{ \VEC{y}}/2)$, where $a$ is the lattice constant.
We are assuming the classical ground state is a cycloidal spin-spiral given by spiral vector $\VEC Q$ along $y$, i.e., with the spins rotating in the $y-z$ plane.
We want to determine which $Q$ correspond to the lowest energy.
Also, we considering only nearest neighbors $J$ and $D$ ($K=0$, $B=0$, $\VEC{D}_{ij} \propto \hat{\VEC{z}} \times \hat{\VEC{r}}_{ij}$). With the help of Fig.~\ref{fig:spinspiralgroundstate}, we have:
\begin{figure}[th!]
  \centering
  \includegraphics[width=0.4\textwidth,trim={0 0 0 0},clip=true]{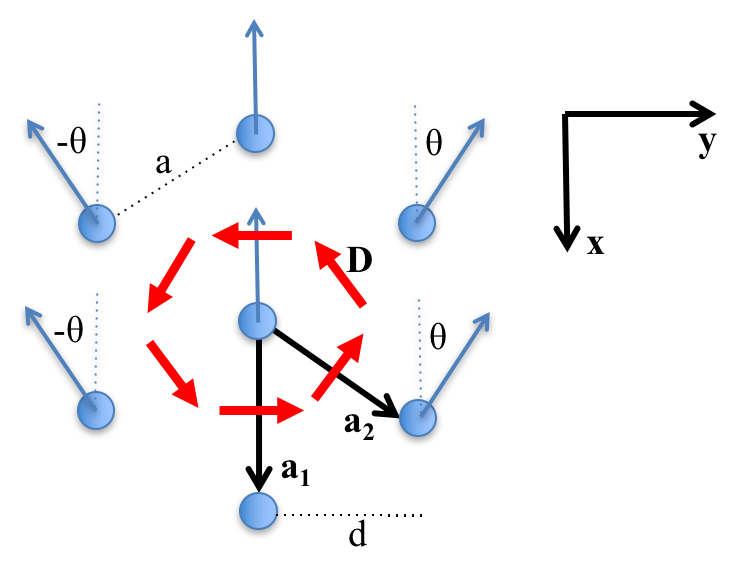}
  \caption{\label{fig:spinspiralgroundstate}
  Sketch of a portion of an hexagonal lattice hosting a cycloidal spin spiral.
  The spins tilt in the $y-z$ plane.
  }
\end{figure}
\begin{equation}
\begin{split}
  \HH = & - \frac{1}{2} \sum_{ij} \left[ J_{ij} \VEC S_i \cdot \VEC S_j + \VEC D_{ij} \cdot (\VEC S_i \times \VEC S_j) \right] \\
   = & - \frac{1}{2} \sum_{ij} \left[ J_{ij} S_i S_j \cos \theta_{ij} + D_{ij}^x  S_i S_j \sin\theta_{ij}  \right] \\
   = & - \frac{1}{2} S^2 \sum_{i} \left[ J(2 + 4\cos \theta ) + 4D^x \sin\theta  \right] \\
   = & - S^2 N \left[ J(1 + 2\cos( d Q ) ) + 2D^x\sin( d Q )  \right] \\
\end{split}
\end{equation}
because $\theta = Q d$, and we have that $d=a\sqrt{3}/2$.
To find the minimal energy, we need to find the zeros of the derivative of this equation in respect to $Q$:
\begin{equation}
  \dv{\HH}{Q} = 2d S^2 N  \left[ J \sin(d Q )  - D^x\cos( d Q )  \right] = 0 \quad ,
\end{equation}
and therefore
\begin{equation}
\begin{split}
  J \sin( d Q) - D^x \cos(d Q)=& 0 \\
  \frac{J}{\sqrt{J^2+{D^x}^2}} \sin( d Q) - \frac{{D^x}^2}{\sqrt{J^2+{D^x}^2}} \cos(d Q)=& 0 \\
  \cos{\alpha}\sin( d Q) - \sin{\alpha}\cos(d Q) =& 0 \\
  \sin( d Q - \alpha)=& 0 \quad , 
\end{split}
\end{equation}
where we defined
\begin{equation}\label{cosalpha}
  \cos \alpha = J/ \sqrt{J^2+{D^x}^2} \eqand \sin \alpha = {D^x}/ \sqrt{J^2+{D^x}^2}  \quad .
\end{equation}
This gives that
\begin{equation}
  \alpha=\arctan({D^x}/J) \quad .  
\end{equation}
For the sine function to be zero its argument has to equal $n\pi$, where $n=0,\pm1,\pm2,...$, which leads to
\begin{equation}
  Q = \frac{n\pi + \alpha}{d} \quad .
\end{equation}
The only two inequivalent solutions are for $n=0,1$.
For all other $n$, a translation by a proper reciprocal lattice vector can bring the solution back to one of these two cases.
If one of the solution is a point of minimal energy the other one has to be of maximal energy.
To check this, we have to take the second derivative of $\HH$:
\begin{equation}
  \dvn{2}{\HH}{Q} = {2d^2 S^2 N} \left\{ J \cos(d Q) + {D^x} \sin(d Q) \right\} \quad ,
\end{equation}
which for the two cases reads (dropping the pre-factor that doesn't matter for the sign analysis):
\begin{equation}
\begin{split}
  n=0 \qquad \qquad \dvn{2}{\HH}{Q} \propto &   J \cos\alpha +{D^x} \sin\alpha  \quad , \\
  n=1 \qquad \qquad \dvn{2}{\HH}{Q} \propto &  -(J \cos\alpha  + {D^x} \sin\alpha)   \quad .
\end{split} 
\end{equation}
This already proves that the two solutions have opposite concavity, therefore one must be a minimum energy point and the other a maximum point.
By using Eq.~\eqref{cosalpha}, we have:
\begin{equation}
\begin{split}
  n=0 \qquad \qquad \dvn{2}{\HH}{Q} \propto & +\frac{J^2+(D^x)^2}{\sqrt{J^2+(D^x)^2}} > 0 \quad , \\
  n=1 \qquad \qquad \dvn{2}{\HH}{Q} \propto & -\frac{J^2+(D^x)^2}{\sqrt{J^2+(D^x)^2}} < 0 \quad ,
\end{split}
\end{equation}
which shows that 
\begin{equation}
  Q=\alpha/d
\end{equation}
is the solution we were looking for.
In the particular case where $J=1$ and $D=2/\sqrt{3}$, such that $D^x=1$, we obtain that $Q=\pi/4d$, which corresponds to a spin-spiral pitch of $\lambda = 8d$.

\subsubsection{Skyrmion lattice} % (fold)
\label{ssub:skyrmion_lattice}

In spherical coordinates $\VEC{S}_{i}$ is uniquely defined by $(\theta_{i}, \phi_{i})$ which represent the polar and azimuthal angles, respectively.
We want to determine the ground state self-consistently.
First, a trial configuration of spins $\VEC{S}_{i}$ is used as a starting point.
Then, we compute the magnetic torques acting on each spin  $\VEC{S}_{i}$: 
\begin{equation}
\begin{split}
\mathcal{T}^{\theta}_{i} = \frac{\partial\mathcal{H}}{\partial\theta_i} \quad , \quad \textnormal{and}  \quad
\mathcal{T}^{\phi}_{i} = \frac{\partial\mathcal{H}}{\partial\phi_i}\quad.
\end{split}
\label{torques}
\end{equation}
The torques $\{\mathcal{T}^{\theta}_{i},\mathcal{T}^{\phi}_{i}\}$ are used to determine the 
set of angles for the next iteration, using a linear mixing:
\begin{equation}
\begin{split}
\theta^{n+1}_{i} = \theta^{n}_{i} + \alpha\,\mathcal{T}^{\theta}_{i}\quad , \quad \textnormal{and}  \quad
\phi^{n+1}_{i}   = \phi^{n}_{i}   + \alpha\,\mathcal{T}^{\theta}_{i}\quad.
\end{split}
\label{mixing_angles}
\end{equation}
$\alpha$ is the mixing parameter, which is set to a small value to ensure convergence.
The output angles from Eq.~\eqref{mixing_angles} are inputted into the Hamiltonian. The torques
are recalculated using Eq.~\eqref{torques}. This process is repeated until self-consistency 
is reached and the magnetic torques acting on each spin $\VEC{S}_{i}$ are zero. 

\begin{figure}[th!]
  \centering
  \includegraphics[width=0.2\textwidth,trim={0 0 0 0},clip=true]{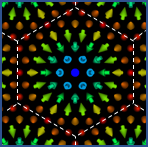}
  \caption{\label{fig:unit_cell_skyrmion}
  The skyrmion lattice ground state. The Hamiltonian parameter were set to $J=1$, $D = J$, $B = 0.36\,J$ and $K = 0.25\,J$. A numerical self-consistent minimization was used to obtain it, as explained in the text.
  }
\end{figure}

We obtained the skyrmion lattice shown in Fig.~\ref{fig:unit_cell_skyrmion} by considering a hexagonal unit cell of 64 atoms.
We chose $J=1$, $D = J$, $B = 0.36\,J$ and $K = 0.25\,J$ as in Ref.~\onlinecite{Roldan2016}.
The direction of the Dzyaloshinskii-Moriya vector is $\hat{\VEC{n}}_{ij} = \hat{\VEC{z}} \times \hat{\VEC{r}}_{ij}$.
We obtained the classical ground state via the numerical minimization approach described above.

\subsection{Holstein-Primakoff transformation} % (fold)
\label{sub:holstein_primakoff_transformation}

Our next step is to determine the excited-states (spin-waves) of our magnetic sample.
We are going to do so in the adiabatic approach, also known as linear spin-wave approximation.
First, we change the frame of reference on each site, so that the z-axis of the new local frame corresponds to the classical spin direction.
Operators in the local frame are indicated by a prime.
This transformation is given by
\begin{equation}\label{refsystrn}
\VEC S_i = \VEC O_i \VEC S'_i \quad ,
\end{equation}
where the rotation matrix is
\begin{equation}\label{eq:rotation_matrix}
\begin{split}
 \VEC O_i =& \, \VEC O^z(\phi_i) \VEC O^y(\theta_i)   \\ =&
\begin{pmatrix}
 \cos\phi_i & -\sin\phi_i & 0\\
 \sin\phi_i &  \cos\phi_i & 0\\
 0          &  0          & 1\\
\end{pmatrix} 
\begin{pmatrix}
 \cos\theta_i & 0            & \sin\theta_i \\
 0            & 1            & 0            \\
-\sin\theta_i & 0            & \cos\theta_i \\
\end{pmatrix} 
\end{split} \quad,
\end{equation}
$\phi_i$ is the polar and $\theta_i$ is the azimuthal angle of the classical ground state orientation of $\VEC S_i$.
Now, we perform a Holstein-Primakoff transformation~\cite{holstein_field_1940,haraldsen_spin_2009,roldan-molina_quantum_2014}, which will replace the spin operator by creation and annihilation spin-wave operators:
\begin{equation}
\VEC S'_i= \VEC M_i \VEC a_i \quad ,
\end{equation}
where
\begin{equation}
\VEC M_i=\sqrt{\frac{S_i}{2}}
\begin{pmatrix}
 1       &       1  & 0 \\
-\MR{i} & \MR{i}  & 0 \\
 0       &       0  & \sqrt{\frac{2}{S_i}} \\
\end{pmatrix} \quad \textnormal{and} \quad
\VEC a_i=
\begin{pmatrix}
{a}_i \\
{a}_i^\dagger \\
{S}_i - a_i^\dagger a_i \\
\end{pmatrix} \eqdot
\end{equation}

Our transformed Hamiltonian is now written as:
\begin{equation}\label{Hamgeneral42}
\begin{split}
\HH=&-\frac{1}{2} \sum_{ij}   \VEC a_i^\dagger \VEC{\tilde{J}}_{ij} \VEC a_j - \sum_i \VEC{\tilde{B}}_i \cdot \VEC a_i \quad ,  
\end{split}
\end{equation}
where
\begin{equation}\label{newJij212}
\renewcommand{\arraystretch}{1.3}
\begin{split}
\VEC{\tilde{J}}_{ij}=&\VEC M_i^\dagger  \VEC O_i^T \VEC J_{ij} \VEC O_j \VEC M_j 
\\ 
= &
\left(\begin{array}{cc|c}
   \tilde J_{ij}^{++}          & \tilde J_{ij}^{+-}           & \tilde J_{ij}^{+z} \\
   \tilde J_{ij}^{-+}          & \tilde J_{ij}^{--}           & \tilde J_{ij}^{-z} \\ \hline
   \tilde J_{ij}^{z+}          & \tilde J_{ij}^{z-}           & \tilde J_{ij}^{zz} \\
\end{array}\right)
=
\begin{pmatrix}
  \VEC A_{ij}^{2\times2}  & \VEC C_{ij}^{2\times1} \\
  \VEC C_{ij}^{1\times2}  & \tilde J_{ij}^{zz} \\
\end{pmatrix}\\
\end{split}
\end{equation}
and
\begin{equation}
\begin{split}
  \VEC{\tilde{B}}_{i}=\VEC B_{i} \VEC O_i \VEC M_i=
  \begin{pmatrix}
    \tilde{B}_i^- & \tilde{B}_i^+ & \tilde{B}_i^z
  \end{pmatrix} \quad .
\end{split}
\end{equation}
We now group terms of different order of the creation/annihilation operators keeping only up to the quadratic order:
\begin{equation}
  \HH = \HH_0 + \HH_1 + \HH_2 \quad ,
\end{equation}
where
\begin{equation} \label{eq:zero_order_hamiltonian}
\begin{split}
  \HH_2 =& -\frac{1}{2} \sum_{ij}   \VEC a_i^\dagger \VEC H_{ij} \VEC a_j    \eqand \\
  \HH_0 =& -\onehalf \tilde J_\VEC{0}^{zz} (\sum_i S_i + N)  - \sum_i \tilde B_i^z (S_i + \onehalf)  \quad ,
\end{split}
\end{equation}
with
\begin{equation}
\begin{split}
  \VEC H_{ij} = &  \VEC A_{ij}^{2\times2} - \big(\tilde{B}_i^z + \tilde J^{zz}_\VEC 0 \big) \VEC I^{2\times2} \delta_{ij}
  = 
  \begin{pmatrix}
    H_{ij}^{++} & H_{ij}^{+-} \\
    H_{ij}^{-+} & H_{ij}^{--} \\
  \end{pmatrix}
    \quad , \\
 \tilde J_\VEC 0^{zz} =& \sum_j \tilde J_{ij}^{zz} S_j \quad \textnormal{and now} \quad
 \VEC a_i =
 \begin{pmatrix}
    a_i \\
    a_i^\dagger
 \end{pmatrix} . 
\end{split}
\end{equation}
The zero-order term $\HH_0$ is a constant and correspond to the energy of the classical ground state.
The first-order $\HH_1$ vanishes if the correct classical ground state has been considered, so we don't list it explicitly.
The second-order $\HH_2$ describes the excited states.

Considering that the system has periodicity given by the translation vectors $\VEC{R}$, we can perform the following Fourier transformation
\begin{equation}\label{eq:fourier_transformation}
    \VEC{a}_{\VEC k} = \frac{1}{\sqrt{N}} \sum_i e^{-\MR{i}\VEC{k} \cdot \VEC{R}_i} \VEC{a}_i \quad , \quad \textnormal{with} \quad
     \VEC{a}_{\VEC k} =
    \begin{pmatrix}
      a_{\VEC k} \\
      a_{-\VEC k}^\dagger \\
    \end{pmatrix} \quad .
\end{equation}
We can then write:
\begin{equation}
   \HH_2 = - \frac{1}{2} \sum_{\VEC k} \VEC a^\dagger_{\VEC k} \VEC H_{\VEC k} \VEC a_{\VEC k}  \quad .
\end{equation}

% subsection holstein_primakoff_transformation (end)

\subsection{Diagonalization and Bogoliubov transformation} % (fold)
\label{sub:bogoliubov_transformation}

To find the spin-wave excitations, we consider the following equation of motion~\cite{erickson_thermodynamics_1991,xiao_theory_2009}:
\begin{equation}\label{eqmot0}
\begin{split}
\MR{i} \dv{\VEC a_i}{t} = [\VEC a_i, \HH_2] \quad .
\end{split}
\end{equation}
By evaluating the commutator in the previous equation, we obtain:
\begin{equation}\label{eq:equation_motion}
\begin{split}
\MR{i} \dv{\VEC a_i}{t} = \sum_j \VEC D_{ij} \VEC a_j \quad ,
\end{split}
\end{equation}
where the dynamical matrix is given by
\begin{equation}\label{eq:dynamical_matrix}
\begin{split}
 \VEC D_{ij} = -\frac{1}{2}
 \begin{pmatrix}
 (H^{++}_{ij} + H^{--}_{ji})  &  (H^{+-}_{ij} + H^{+-}_{ji}) \\
-(H^{-+}_{ij} + H^{-+}_{ji})  & -(H^{++}_{ji} + H^{--}_{ij}) \\
\end{pmatrix} \quad .
\end{split}
\end{equation}
Because $\HH_2$ is Hermitian, the following relations hold:
\begin{equation}
\begin{split}
H^{++}_{ij} = H^{--}_{ji}  \quad , \quad  H^{++}_{ij} = (H^{--}_{ij})^*  \quad , \\  H^{+-}_{ij} = H^{+-}_{ji} \quad , \quad   H^{+-}_{ij} = (H^{-+}_{ij})^* \quad .
\end{split}
\end{equation} 
Therefore, the dynamical matrix in Eq.~\eqref{eq:dynamical_matrix} can be simplified to
\begin{equation}
\begin{split}
 \VEC D_{ij} =
 \begin{pmatrix}
-H^{++}_{ij}  & -H^{+-}_{ij} \\
H^{-+}_{ij}   &  H^{--}_{ij} \\
\end{pmatrix}
=
\VEC g \VEC H_{ij} \quad,
\end{split}
\end{equation}
where
\begin{equation}
  \VEC g = 
  \begin{pmatrix}
    -1 & 0 \\
     0 & 1 \\
  \end{pmatrix} \quad.
\end{equation}
The matrix $\VEC{g}$ embodies the commutation relations of the Holstein-Primakoff bosons.
Note that $\VEC g \VEC g = \VEC 1$.
Considering the Fourier transformation of Eq.~\eqref{eq:fourier_transformation} and assuming stationary solutions of these operators $\VEC a_\VEC k$, such that they depend on time only via a global phase, as in
\begin{equation}
  \VEC a_{\VEC k}(t) = e^{-\MR{i}\omega_\VEC k t}\VEC a_{\VEC k} \quad ,
\end{equation}
we obtain for Eq.~\eqref{eq:equation_motion} the following eigenvalue equation:
\begin{equation}\label{ultimate}
\begin{split} 
 \VEC D_{\VEC k} \VEC a_{\VEC k}  = \omega_\VEC k \VEC a_{\VEC k} \quad .
\end{split}
\end{equation}

For the general problem, we diagonalize $\VEC D_{\VEC k}$ numerically, but for some simple cases it can also be solved analytically.
We now show how diagonalizing $\VEC D_{\VEC k}$ provides the eigenvalues and eigenfunctions of $\VEC H_{\VEC k}$.
For simplicity, we are going to drop the $\VEC k$ index.
$\VEC D$ is not Hermitian, therefore we need to define left and right eigen-solutions as follows:
\begin{equation}
  \VEC D \mathcal{R}_{r} = \omega_r \mathcal{R}_r  \quad , \quad
  \mathcal{L}_r \VEC D   = \omega_r \mathcal{L}_r  \quad ,
\end{equation}
where $\mathcal R_r$ is a column eigenvector, $\mathcal L_r$ is a row eigenvector and $r$ is the eigenvalue index.
In matrix form this can be written as
\begin{equation}
  \VEC D \bm{\mathcal{R}} = \bm{\mathcal{R}} \VEC\Omega \quad , \quad
  \bm{\mathcal{L}} \VEC D = \VEC\Omega \bm{\mathcal{L}} \quad ,
\end{equation}
where $\bm{\mathcal{L}}$ and $\bm{\mathcal{R}}$ contain all left and right eigenvectors of $\VEC D$ as rows and columns, respectively.
$\bm\Omega$ is a diagonal matrix containing the eigenvalues of $\VEC D$.
In this way, we have that
\begin{equation} \label{eq:omega}
  \bm{\mathcal{L}} \VEC D \bm{\mathcal{R}} = \bm{\mathcal{L}} \bm{\mathcal{R}} \VEC\Omega \quad .
\end{equation}
Because we want that $\bm{\mathcal{R}}$ to represent boson operators, it must satisfy the proper commutation relations that can be expressed as~\cite{erickson_thermodynamics_1991,toth_linear_2015}:
\begin{equation} \label{eq:boson_commutation_relation}
  \bm{\mathcal{R}}^\dagger \VEC g \bm{\mathcal{R}} = \VEC g \quad , \quad \bm{\mathcal{R}} \VEC g \bm{\mathcal{R}}^\dagger = \VEC g \quad .
\end{equation}
Based on Eq.~\eqref{eq:boson_commutation_relation}, we can show that knowing the right eigenvectors we can construct the left ones via:
\begin{equation}
  \bm{\mathcal{L}} = \VEC g \bm{\mathcal{R}}^\dagger \VEC g \quad ,
\end{equation}
which implicates in $\bm{\mathcal{L}} \bm{\mathcal{R}} = \bm{\mathcal{R}} \bm{\mathcal{L}} = \VEC 1$.
Here comes the proof:
\begin{equation}
\begin{split}
  \VEC D \bm{\mathcal{R}} =& \bm{\mathcal{R}} \VEC\Omega \\
 \VEC g\VEC g \VEC D \bm{\mathcal{R}} =& \bm{\mathcal{R}} \VEC \Omega \\
 \bm{\mathcal{R}} \VEC g \bm{\mathcal{R}}^\dagger \VEC g \VEC D \bm{\mathcal{R}} =& \bm{\mathcal{R}} \VEC \Omega \\
 \VEC g \bm{\mathcal{R}}^\dagger \VEC g \VEC D \bm{\mathcal{R}} =& \VEC \Omega \\
 \bm{\mathcal{L}} \VEC D \bm{\mathcal{R}} =& \VEC \Omega \\
 \bm{\mathcal{L}} \VEC D \bm{\mathcal{R}} \bm{\mathcal{L}} =& \VEC \Omega \bm{\mathcal{L}} \\
 \bm{\mathcal{L}} \VEC D =& \VEC \Omega \bm{\mathcal{L}} \quad .
\end{split}
\end{equation}

Starting from Eq.~\eqref{eq:omega}, we have:
\begin{equation}\label{ultimate2}
\begin{split}
\bm{\mathcal{L}} \VEC D \bm{\mathcal{R}} &= \VEC \Omega \\
\VEC g \bm{\mathcal{R}}^\dagger \VEC g \VEC D \bm{\mathcal{R}} &= \VEC \Omega \\
\bm{\mathcal{R}}^\dagger \VEC H \bm{\mathcal{R}} &= \VEC g \VEC \Omega = \VEC \Lambda \\
\VEC H &= \bm{\mathcal{L}}^\dagger \VEC \Lambda \bm{\mathcal{L}} \quad , \\
\end{split}
\end{equation}
where $\VEC \Lambda = \VEC g \VEC \Omega$ is diagonal and positive.
This equation reveals that $\bm{\mathcal{R}}$ generates a transformation into a basis where the Hamiltonian is diagonal:
\begin{equation} 
\begin{split}
        \HH_2 = & - \frac{1}{2} \sum_{\VEC k} \VEC a^\dagger_{\VEC k} \VEC H_{\VEC k} \VEC a_{\VEC k} 
         = - \frac{1}{2} \sum_{\VEC k} \VEC a^\dagger_{\VEC k} \bm{\mathcal{L}}_{\VEC k}^\dagger \VEC \Lambda_{\VEC k} \bm{\mathcal{L}}_{\VEC k} \VEC a_{\VEC k} \\
         = & - \frac{1}{2} \sum_{\VEC k} \VEC b^\dagger_{\VEC k} \VEC \Lambda_{\VEC k} \VEC b_{\VEC k}   \quad , 
\end{split}
 \end{equation} 
\begin{equation}
\begin{split}
   \HH_2 = & - \frac{1}{2} \sum_{\VEC k} \VEC b^\dagger_{\VEC k} \VEC \Lambda_{\VEC k} \VEC b_{\VEC k} 
         = - \frac{1}{2} \sum_{\VEC k} \VEC b^\dagger_{\VEC k} \bm{\mathcal{R}}^\dagger_{\VEC k} \VEC H_{\VEC k} \bm{\mathcal{R}}_{\VEC k}  \VEC b_{\VEC k} \\
         = & - \frac{1}{2} \sum_{\VEC k} \VEC a^\dagger_{\VEC k} \VEC H_{\VEC k} \VEC a_{\VEC k}  \quad , 
\end{split}
 \end{equation} 
 where
 \begin{equation}\label{eq:bogoliubov_transformation}
 \begin{split}
  \VEC b_{\VEC k} =& \bm{\mathcal{L}}_{\VEC k} \VEC a_{\VEC k} \quad , \quad
  \VEC b^\dagger_{\VEC k} =\VEC a^\dagger_{\VEC k} \bm{\mathcal{L}}_{\VEC k}^\dagger \quad , \\
  \VEC a_{\VEC k} =& \bm{\mathcal{R}}_{\VEC k} \VEC b_{\VEC k} \quad , \quad
  \VEC a^\dagger_{\VEC k} =  \VEC b^\dagger_{\VEC k} \bm{\mathcal{R}}_{\VEC k}^\dagger \quad .
 \end{split}
 \end{equation}
$\VEC b_{\VEC k}$ and $\VEC b_{\VEC k}^\dagger$ are a new set of boson annihilation and creation operators.
This transformation is known as the Bogoliubov transformation~\cite{erickson_thermodynamics_1991,xiao_theory_2009,haraldsen_spin_2009,roldan-molina_quantum_2014,toth_linear_2015}.

% subsection bogoliubov_transformation (end)

\subsection{Spin waves modes in: a ferromagnet, a spin-spiral and a skyrmion lattice} % (fold)
\label{sub:spin_waves_modes}

\begin{figure}[th!]
  \centering
  \includegraphics[width=0.45\textwidth,trim={0 50 40 0},clip=true]{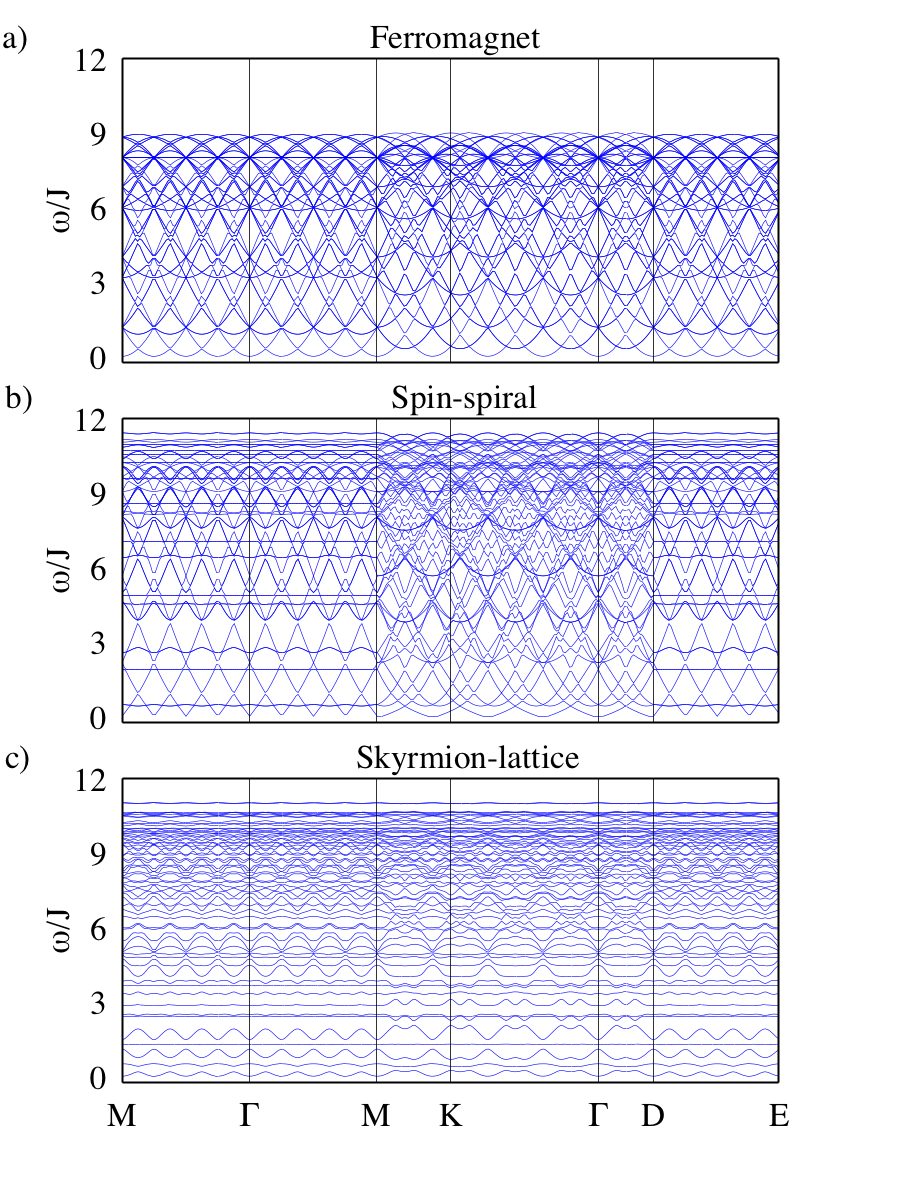}
  \includegraphics[width=0.15\textwidth,trim={0 0 0 0},clip=true]{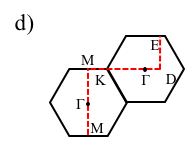}
  \caption{\label{fig:dispersion}
  \textbf{Spin-wave dispersion relations of non-collinear magnets.}
  The dispersion is composed by the eigenvalues given by Eq.~\ref{ultimate2}.
  Panels (a), (b) and (c) correpond to a ferromagnet, a spin-spiral and a skyrmion lattice in a hexagonal monolayer.
  Each of these phases contained 64 atoms in the unit cell, and they shared the same primitive lattice.
  The dispersions were calculated through the reciprocal path shown in (d).
  We considered a polarization along $z$, e.g. normal to the monolayer plane.
  Parameters: ferromagnet \{$J=1$\}; spin-spiral \{$J=1$, $D=2J/\sqrt{3}$\}; skyrmion lattice \{$J=1$, $D=J$, $K=0.25J$, $B=0.36J$\}.
  }
\end{figure}

Now, we would like to show the dispersion-relation obtained with the formalism of the previous sections for a ferromagnet, a spin-spiral and a skyrmion lattice.
The dispersion-relation consist of the eigenvalues of the Hamiltonian as a function of the wavevector $\VEC k$, which evolves solving Eqs.~\ref{ultimate} and ~\ref{ultimate2}. 
We calculated all these three cases with the same hexagonal Bravais lattice of primitive vectors  $\VEC a_1= a \hat{\VEC x}$ and $\VEC a_2= a(\frac{1}{2}\hat{\VEC x} + \frac{\sqrt{3}}{2} \hat{\VEC y})$, with $a=8$; and with same unit cell containing 64 atoms.
The ground state spin configurations inputted were the one obtained in Sec.~\ref{sec:ground_state}.
The same parameters as used for the ground-state determination were used: ferromagnet \{$J=1$\}; spin-spiral \{$J=1$, $D=2J/\sqrt{3}$\}; and skyrmion lattice \{$J=1$, $D=J$, $K=0.25J$, $B=0.36J$\}.

Fig.~\ref{fig:dispersion}(a) shows the dispersion curves for the ferromagnetic case.
Many bands appear because the Goldstone mode of the ferromagnet gets folded due to the reduction of the Brillouin zone when considering many atoms in the unit cell.
Fig.~\ref{fig:dispersion}(b-c) present the dispersion curves for the spin-spiral and the skyrmion lattice.
We can observe that the dispersion curves of the skyrmion lattice feature many gaps and some dispersionless bands.
The dispersion relations were calculated through the reciprocal space path shown in Fig.~\ref{fig:dispersion}(d).

The motion of the atomic spin moments corresponding to the spin-wave modes can be seen on the videos in the Supplementary Materials.
Videos 1, 2 and 3 represent the lowest-energy excitations of the spin-spiral sample at $-$Q, $\Gamma$ and $+$Q, respectively.
Meanwhile, videos 4 to 8 display the dynamics of the five lowest-energy spin-waves of the skyrmion lattice at the $\Gamma$-point.
The central gray arrow represents the total atomic spin.
Also, the amplitude of the precession of the local spin were rescaled to enhance the motion.
We used the following equations to describe the spin precession of every site in the local reference frame:
\begin{equation}
\begin{split}
  {S'}^{x,r}_i(\VEC k) = & A^{x,r}_i \cos( \omega_r t + \VEC{R}_i \cdot \VEC k + \phi_i^{x,r}) \quad , \\
  {S'}^{y,r}_i(\VEC k) = & A^{x,r}_i \sin( \omega_r t + \VEC{R}_i \cdot \VEC k + \phi_i^{y,r}) \quad , \\
  {S'}^{z,r}_i(\VEC k) = & 1 \quad ,
\end{split}
\end{equation}
where the phases and amplitudes were obtained from the calculated right-eigenvectors via:
\begin{equation}
\begin{split}
  \RR^{+,r}_{i} = A^{x,r}_i e^{\MR{i} \phi_i^{x,r}} \quad \textnormal{and} \quad \RR^{-,r}_{i} = A^{y,r}_i e^{\MR{i} \phi_i^{y,r}} \quad .
\end{split}
\end{equation}
Here, $i$ labels the atomic sites, $r$ is the mode index and $\VEC k$ the wavevector of the spin-wave. $\RR^{r}_{i}$ are the right-eigenvector elements. 
Then, the precessing spin were brought into the global reference frame via:
\begin{equation}
  \VEC S_i^r= \VEC O_i \VEC{S'}_i^r \quad .
\end{equation}

% subsection spin_waves_modes (end)
% section adiabatic_approach_of_spin_waves_for_noncolinear_systems (end)

\subsection{Spin-spin correlation tensor for non-collinear magnets} % (fold)
\label{sub:spin_spin_correlation_tensor_for_non_collinear_magnets}

To understand what out of the many spin-wave bands from Fig.~\ref{fig:dispersion} can be actually excited and detected with inelastic electron scattering, we need to calculate the scattering rate given by Eq.~\eqref{eq:scattering_rate}.
We start by defining a spin excitation $\ket{\VEC{k} r}$ of wavevector $\VEC k$ and mode index $r$ as created by the action of a new boson operator on a new ground-state $\ket{\tilde 0}$: 
\begin{equation}\label{eq:operators_action}
    b_{r}^\dagger(\VEC k) \ket{\tilde0}  = \ket{\VEC{k} r}  \quad , \quad
    b_{r}(\VEC k) \ket{\tilde0} = 0 \quad  \textnormal{and} \quad 
    \braket{\VEC{k} r|\tilde0} = 0 \quad ,
\end{equation}
such that
\begin{equation}
  \HH_2 \ket{\VEC{k} r} = \omega_r(\VEC k) \ket{\VEC{k} r} \quad .
\end{equation}

Also, the relation between the new and the old boson operators from Eq.~\eqref{eq:bogoliubov_transformation} can be rewritten as:
\begin{equation} \label{eq:bogoliubov_transformation2}
  a_\mu^\alpha(\VEC k) = \sum_{\beta,r} \mathcal{R}_{\mu r}^{\alpha \beta}(\VEC k) b_{r}^\beta(\VEC k) \quad ,
\end{equation}
where $\alpha,\beta=\pm$ to represents the creation or annihilation operators ($a^+=a^\dagger$ and $a^-=a$); and $\mu,\nu$ are site indexes within a unit cell.
We can now see that the classical ground-state $\ket0$ is not annhilated by $b_{r}(\VEC k)$, because it is a combination of $a_\mu(\VEC k)$ and $a^\dagger_\mu(\VEC k)$.
This leads to the definition of a modified ground-state in Eq.~\ref{eq:operators_action}.

For the scattering rate, we need to evaluate the spin-spin correlation tensor of Eq.~\ref{eq:spinspin_tensor}:
\begin{equation}\label{eq:density_tensor}
\begin{split}
  \mathcal{N}_{\mu\nu}^{\alpha\beta}(\VEC q, \omega) =
   \sum_{\VEC{k} r} \delta\left(\omega-\omega_{r}(\VEC k)\right)
   \braket{\tilde 0|S_{\mu}^\alpha(\VEC q)|{\VEC{k} r}} & \times \\
   \braket{{\VEC{k} r}|S_{\nu}^\beta(\VEC q)|\tilde 0} & \quad,
\end{split}
\end{equation}
where 
\begin{equation}
  S_{\nu}^\beta(\VEC q) =  \frac{1}{\sqrt{N_l}} \sum_n e^{\MR{i} \VEC{q} \cdot \VEC{R}_n} S_{n \nu}^\beta  \quad .
\end{equation}
We can rewrite the spin operator as:
\begin{equation}
\begin{split}
  S_{n \nu}^\beta 
  = & O_\nu^{\beta +}{S'}_{n\nu}^+  + O_\nu^{\beta -}{S'}_{n\nu}^- + O_\nu^{\beta z}{S'}_{n\nu}^z \\
  = & O_\nu^{\beta +} \sqrt{2 S_\nu} {a}_{n\nu}  + O_\nu^{\beta -}\sqrt{2 S_\nu} {a}_{n\nu}^\dagger + \\
    & +  O_\nu^{\beta z}( S_{n\nu} - a_{n\nu}^\dagger a_{n\nu}) \quad ,
\end{split}
\end{equation}
where $S'^\alpha$ is the spin operator in the local reference frame related to the global representation via the rotation matrix $O_\nu^{\alpha \beta}$.
We obtain that the left matrix element in Eq.~\eqref{eq:density_tensor} reads 
\begin{equation} \label{eq:correlation_function}
\begin{split}
    &\braket{\tilde0|S_{\mu}^\alpha(\VEC{q})|{\VEC{k}, r}} 
    =  \frac{1}{\sqrt{N_l}} \sum_n e^{\MR{i} \VEC{q} \cdot \VEC{R}_n} \times \\
    & \left(
    O_\mu^{\alpha +}  \sqrt{2 S_\mu}   \braket{\tilde 0| a_{n\mu}| \VEC k, r} + \right. \\
  &  \left. +   O_\mu^{\alpha -}  \sqrt{2 S_\mu}   \braket{\tilde 0| a_{n\mu}^\dagger|{\VEC{k}, r} } 
    + O_\mu^{\alpha z} \braket{\tilde 0| a_{n\mu}^\dagger a_{n\mu} |{\VEC{k}, r} } \right) \quad .
\end{split}
\end{equation}
Using Eqs.~\eqref{eq:operators_action} and \eqref{eq:bogoliubov_transformation2}, and the boson commutation relations, the RHS terms of the previous equation are then given by
\begin{equation}
\begin{split}
    \braket{\tilde 0|a_{n\mu}|\VEC{k}, r}  =&   \frac{1}{\sqrt{N_l}} e^{\MR{i} \VEC{k} \cdot \VEC{R}_n} \mathcal{R}_{\mu r}^{--} \\
    \braket{ \tilde 0|a_{n\mu}^\dagger|\VEC{k}, r} =&   \frac{1}{\sqrt{N_l}} e^{\MR{i}\VEC{k} \cdot \VEC{R}_n} \mathcal{R}_{\mu r}^{+-}  \\
    \braket{\tilde 0|a_{n\mu}^\dagger a_{n\mu}|\VEC{k}, r} =&  0 \eqdot
\end{split}
\end{equation}
Then, Eq.~\eqref{eq:correlation_function} becomes:
\begin{equation}
\begin{split}
    \braket{\tilde0|S_{\mu}^\alpha(\VEC{q})|{\VEC{k}, r}} =  &   
     \sqrt{2 S_\mu} \delta(\VEC{q} + \VEC{k}) \times \\
     &
     \left( 
     O_\mu^{\alpha +}  \RR_{\mu r}^{--}(\VEC{k})   +
     O_\mu^{\alpha -}  \RR_{\mu r}^{+-}(\VEC{k}) 
     \right)  
\end{split}
\end{equation}
where we used $ \frac{1}{{N_l}} \sum_n e^{\MR{i} (\VEC{q} - \VEC{k}) \cdot \VEC{R}_n} = \delta(\VEC{q} - \VEC{k}) $.
In a similar way, we obtain the right matrix element in Eq.~\eqref{eq:density_tensor}:
\begin{equation}
\begin{split}
    \braket{{\VEC{k}, r}|S_{\nu}^\beta(\VEC q)|\tilde0} 
    =& \sqrt{2 S_\nu} \delta(\VEC{q} - \VEC{k}) \times \\
     & \left( 
     O_\nu^{\beta +}  \RR_{\nu r}^{-+}(\VEC{k})   +
     O_\nu^{\beta -}  \RR_{\nu r}^{++}(\VEC{k}) 
     \right)  \quad .
\end{split}
\end{equation}
Plugging back to Eq.~\eqref{eq:density_tensor}, we have our final expression:
\begin{equation}\label{eq:density_tensor2}
\begin{split}
    \NN_{\mu\nu}^{\alpha\beta}(\VEC{q}, \omega) = &  2\sqrt{ S_\mu S_\nu} \sum_{r} \delta\big(\omega-\omega_{r}(\VEC{q})\big)  \times  \\
    &  \left[
         O_\mu^{\alpha +}  (\RR_{\mu r}^{++}(\VEC{q}))^*   +
         O_\mu^{\alpha -}  (\RR_{\mu r}^{-+}(\VEC{q}))^*
       \right] \times \\
    &  \left[
         O_\nu^{\beta +}  \RR_{\nu r}^{-+}( \VEC{q})   +
         O_\nu^{\beta -}  \RR_{\nu r}^{++}( \VEC{q})  
       \right] \quad .
\end{split}
\end{equation}

Also, we need to be able to transform the rotation matrix from the $xyz$ representation into the $+-z$.
This is given by
\begin{equation}
  \VEC O^+ =  \VEC M' \VEC O \VEC M'^{-1} \quad ,
\end{equation}
where
\begin{equation}
\VEC M' =
\begin{pmatrix}
1 & \MR{i} & 0 \\
1 &-\MR{i} & 0 \\
0 & 0 & 1 \\
\end{pmatrix} \quad .
\end{equation}

\subsubsection{Spin-resolved spectra (SREELS): Ferromagnet} % (fold)
\label{fig:ferromagnetic_spectra}

\begin{figure*}[th!]
  \centering
  \includegraphics[width=0.7\textwidth,trim={5 50 0 0},clip=true]{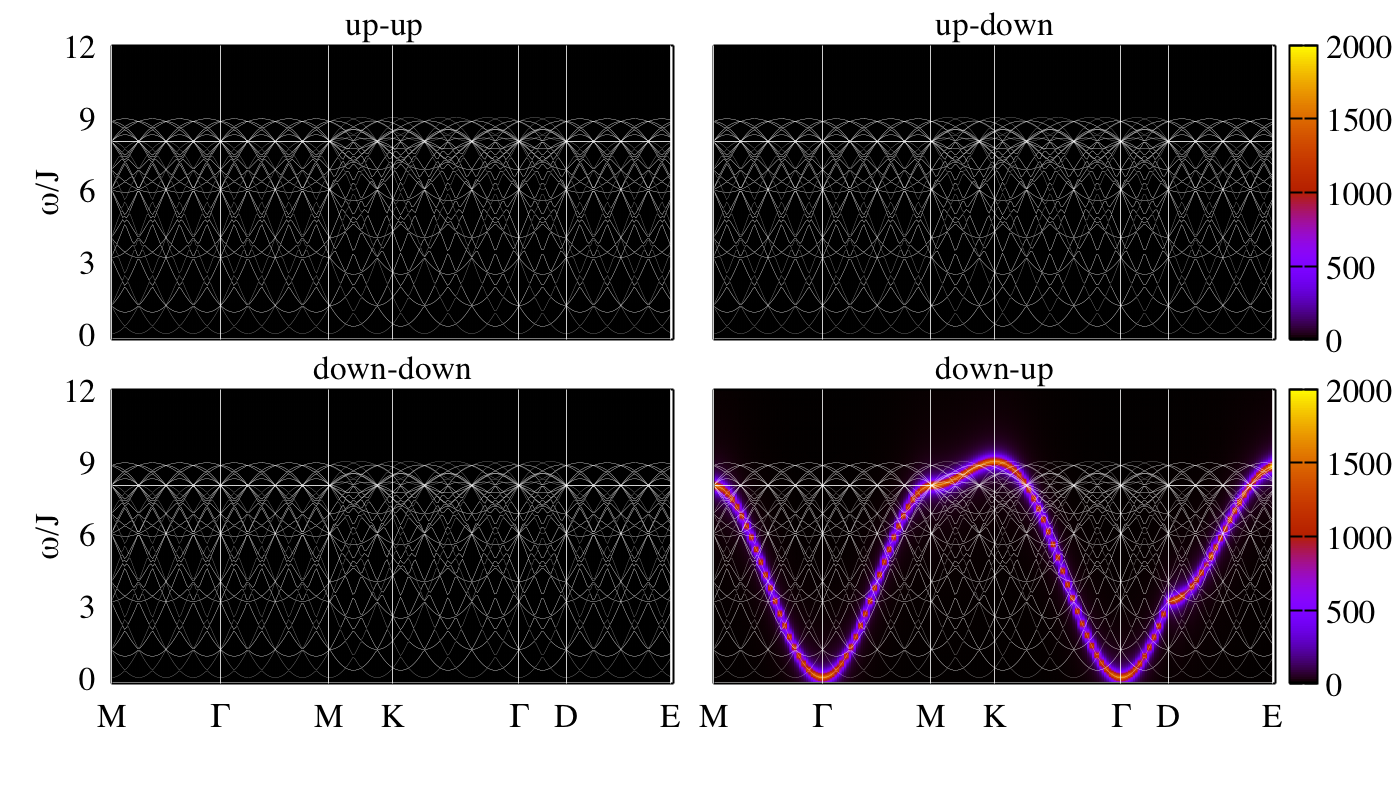}
  \caption{\label{fig:dispersionsFMz}
  \textbf{Spin-resolved spin-wave spectra in a ferromagnet.}
  The system consists of a hexagonal ferromagnetic monolayer with 64 atoms in the unit cell, as in Sec.~\ref{sub:spin_waves_modes}.
  The inelastic spectra are given by the color maps, and they were obtained from Eq.~\eqref{eq:scattering_rate} with the help of Eq.~\ref{eq:density_tensor2}.
  Meanwhile, the gray lines represent the many dispersion-relation curves of the spin-waves, see Fig.~\ref{fig:dispersion}(a).
  The electron beam polarization is aligned along $z$.
  On the model, only a uniform nearest neighbour $J$ was considered.
  }
\end{figure*}

Fig.~\ref{fig:dispersionsFMz} shows the SREELS spectra for different spin-channels of a ferromagnet hexagonal monolayer, with magnetization along $z$, discussed in Sec.~\ref{sub:spin_waves_modes}.
The dispersion curves are shown by gray lines.
The polarization was taken along the precession axis of the spin-waves.
Only one channel responds, revealing the Goldstone mode of ferromagnet.

%***********************************************************************
\bibliography{bibliography}

\end{document}